\documentclass{aa}  

\usepackage{graphicx}
\usepackage{txfonts}
\usepackage{color}
\usepackage[normalem]{ulem}

\begin{document}

   \title{
   Heavy-element Rydberg transition line emission from the post-giant-evolution star HD\,101584
   }

   \titlerunning{Heavy-element Rydberg transition lines towards HD\,101584}

   \author{H.~Olofsson  \inst{1}
          \and 
          J.H.~Black    \inst{1}
          \and
          T.~Khouri    \inst{1}
          \and
          W.H.T.~Vlemmings \inst{1}
          \and      
          E.M.L.~Humphreys \inst{2}
          \and
          M.~Lindqvist \inst{1}
          \and
          M.~Maercker \inst{1}
          \and
          L.~Nyman \inst{3}
          \and
          S.~Ramstedt \inst{4}
          \and
          D.~Tafoya  \inst{1}          
}

   \institute{Dept of Space, Earth and Environment, Chalmers Univ. of Technology,
              Onsala Space Observatory, SE-43992 Onsala, Sweden\\
              \email{hans.olofsson@chalmers.se}
         \and
         ESO, Karl-Schwarzschild-Str. 2, D-85748 Garching bei M{\"u}nchen, Germany
         \and
         ESO, Alonso de Cordova 3107, Vitacura, Santiago, Chile 
         \and
         Dept of Physics and Astronomy, Uppsala University, Box 516, SE-75120 Uppsala, Sweden              
}

   \date{Received 12 February 2021 / Accepted 20 April 2021}

 \abstract{}{}{}{}{} 
 
  \abstract
   {We report the detection of two lines at millimetre wavelengths towards the immediate surroundings of the post-giant and most likely post-common-envelope star HD\,101584 using high-angular-resolution ALMA observations. The circumstellar environment of this object is rich in different molecular species, but we find no viable identifications in terms of molecular lines.}
   {We aim to determine whether or not these lines can be attributed to the Rydberg transitions ---X30$\alpha$ and X26$\alpha$--- of neutral atoms of elements heavier than carbon.}
   {A simple model in strict local thermodynamic equilibrium for a warm-gas environment of the moderate-temperature star ($T_{\rm eff}$\,$\approx$\,8500\,K) was constructed to corroborate our findings. A geometrically thin, disc-like geometry seen face-on was chosen and a distance of 1\,kpc.}
  {The observed flux densities of the lines and the continuum at 232 and 354\,GHz can be reproduced using $\approx$\,10$^{-3}$\,$M_\odot$ of gas at a temperature of $\approx$\,2800\,K and a hydrogen density of $\approx$\,10$^{12}$\,cm$^{-3}$, assuming solar abundances for the elements. The gas lies within a distance of about 5\,au from the star (assuming a distance of 1\,kpc). The ionisation fraction is low, $\approx$\,3$\times$10$^{-5}$. The origin of such a region is not clear, but it may be related to a common-envelope-evolution phase. With these conditions, the line emissions are dominated by Rydberg transitions within the stable isotopes of Mg.  A turbulent velocity field in the range 5.5\,--\,7.5\,km\,s$^{-1}$ is required to fit the Gaussian line shapes. An upper limit to the average magnetic field in the line-emitting region of 1\,G is set using the Zeeman effect in these lines.}
   {We speculate that Rydberg transitions of heavy elements may be an interesting probe for the close-in environments of other moderate-temperature objects like AGB stars, red supergiants, yellow hypergiants, and binaries of various types.}

   \keywords{circumstellar matter --
          Stars: individual: HD101584 --
          Stars: AGB and post-AGB -- 
          Radio lines: stars
               }

   \maketitle
%
%
%
\section{Introduction}

As pointed out already by \citet{bohr20}, in the limit of large quantum numbers $n,$ all atomic spectra appear `hydrogenic' with the ionisation energy of a single electron far away from the ionic core well described by the classical Rydberg formula, $E_{\rm n}$\,$\propto$\,$n^{-2}$. Such highly excited atoms occur frequently in different astronomical objects where conditions are favourable; for instance, at low densities and in intense photo-ionising fields. The theory of Rydberg transitions in the cosmic nebular environment was first developed by Menzel and collaborators in a series of papers starting in 1937 \citep[e.g.][]{menz37}. The possibility to detect such cosmic lines at radio wavelengths was first discussed by \citet{vandhu45} and clarified quantitatively by \citet{kard59}, specifically in the tenuous nebular medium which is exposed to intense UV light, where highly excited Rydberg states are populated by radiative recombination of the parent ion and a free electron; hence the name, (radio) `recombination' lines. 

The first detections of cosmic radio recombination lines from hydrogen were made by \citet{dravdrav64} and \citet{hoglmezg65} towards H\,II regions. Subsequently, recombination lines from helium were detected towards H\,II regions, and carbon lines were detected from a cooler medium surrounding H\,II regions \citep{lilletal66, panketal77}. \citet{chaietal72} found a weak feature on the blueshifted side of the C158$\alpha$ line, and identified it as due to one or a combination of the heavy element species Mg, Si, S, and Fe. Safe identification of the carrier is often hampered by the fact that the line separations ---only due to the differences in reduced mass--- of elements heavier than carbon are smaller than the line widths. However, in the cold $\rho$\,Oph dark cloud, \citet{chai75} managed to identify a sulphur recombination line. Recombination lines have now been detected also towards other ionising sources like planetary nebulae \citep{roeletal91, bachetal92, sacoetal17}, luminous blue variables like \object{$\eta$~Car} \citep{coxetal95}, and young stellar objects \citep{mapietal89, thumetal13}. Comprehensive reviews of recombination lines are given in \citet{browetal78} and \citet{gordsoro02}. Also the Sun has been detected in Rydberg transitions at submillimetre (submm; H and possibly Mg \citep{claretal00a, claretal00b}) and mid-infrared (Mg and Al \citep{murcetal81, braunoye83, channoye83}) wavelengths. Here the conditions are different, and a different approach is required to describe the excitation \citep{carlrutt92}; therefore, using the nomenclature of Rydberg transition lines (RTLs) rather than recombination lines is more appropriate. 

Here we present the detection of spectral lines at about 232 and 354\,GHz towards the moderate-temperature, low-mass star \object{HD\,101584} using the Atacama Large Millimeter/submillimeter Array (ALMA). We argue that these lines are RTLs of neutral atoms of elements heavier than carbon. A simple model in strict local thermal equilibrium (LTE) is used to estimate the viability of this proposal, and to estimate the physical characteristics of the line-emitting region and identify the most likely carrier (or carriers) of the lines. 

%
%
%
%
\section{HD\,101584}
\label{s:hd101584}

HD\,101584 is a binary system with a significant infrared excess. For the primary star, we adopt a spectral type of A6Ia and effective temperature of 8500\,K; see \citet{kipp05} who determined the temperature from spectral synthesis of high-resolution optical spectra. It has been difficult to establish whether its current state represents evolution beyond the asymptotic giant branch (AGB) or the red giant branch (RGB). It was originally classified as a post-AGB star \citep{partpott86, bakketal96a}, but \citet{olofetal19} classified it as a post-RGB star, the consequence of premature termination of the RGB caused by common-envelope (CE) evolution that ended before the two stars merged. 

\citet{olofetal19} studied the circumstellar environment of HD\,101584 in detail using ALMA. This object is morphologically and kinematically complex,  consisting of a bipolar high-velocity outflow (seen almost pole-on) surrounded by an hourglass structure, an equatorial density enhancement (seen almost face-on), and a central compact source (whose geometry and orientation are unknown). A total of 12 different molecular species (not counting isotopologues) have been identified in the circumstellar medium. Here we recapitulate our present knowledge of the near environment of HD\,101584. For more information on this source, see \citet{olofetal19} and references therein.

Many of the molecular line emissions observed using ALMA show a compact component that coincides in space with the position of HD\,101584, the central compact source (CCS) in the nomenclature of \citet{olofetal19}; detected molecular species are CO, SiO, SiS, CS, SO, SO$_2$, OCS, and H$_2$S. The emission from this component is partially resolved, and its estimated mass is $\approx$\,0.03\,[$D$/1\,kpc]$^2$\,$M_\odot$, which corresponds to $\approx$\,7\,\% of the total circumstellar gas mass. The angular size varies with the observed molecular transition, but lies in the range 0\farcs1--0\farcs15 (FWHM of Gaussian fit to the brightness distribution deconvolved with the synthesised beam), that is, within $\approx$\,75\,[$D$/1\,kpc]\,au of HD101584. The morphology of this component has not been determined, except for the innermost structure observed with the VLTI/PIONIER instrument. \citet{klusetal20} detected an almost circular structure interpreted to be the sublimation rim of a circumbinary dust disc seen close to face on. The radius of this ring is about 2\,[$D$/1\,kpc]\,au. The lines discussed in this paper most likely come from a region within a radius of about 5\,[$D$/1\,kpc]\,au from the binary. The binary separation is estimated to be about 0.5\,au. 

The systemic velocity of HD\,101584 is estimated to be 41.5\,$\pm$\,0.15\,km\,s$^{-1}$ (in LSR scale). We have derived the average velocity from Gaussian fits to six ALMA lines that show strong emission from the CCS: CO(\mbox{2--1}), C$^{18}$O(\mbox{2--1}), SO(\mbox{$5_5-4_4$}), SO(\mbox{$5_6-4_5$}), SO$_2$(\mbox{$16_{3,13}-16_{2,14}$}), and H$_2$S(\mbox{$2_{20}-2_{11}$}); see \citet{olofetal19}. The systemic velocity is crucial for deriving the rest frequencies of the lines discussed in this paper.

The parallax of HD\,101584 in the second Gaia data release suggests a distance as large as 2\,kpc \citep{gaiaetal16, gaiaetal18}. The third Gaia data release maintains this distance, but the excess noise required in the fitting process has increased \citep{gaiaetal20}. However, as argued by \citet{olofetal19} there are reasons to suspect that this distance is not correct, the primary one being that the estimated angular distance between HD\,101584 and its companion is comparable to the parallax. Consequently, we follow their adopted procedure of presenting all results for a nominal distance of $D$\,=\,1.0\,kpc. From a fit to the spectral energy distribution, \citet{olofetal19} derived a luminosity and radius of the star of 1600\,$L_\odot$ and 19\,$R_\odot$, respectively, at 1\,kpc, based on the assumption that the radiation of the infrared excess is not isotropic.

%
%
%
\section{Observations}

\subsection{ALMA data}
\label{s:alma_obs_desc}

The ALMA data were obtained during cycles~1 (May 2014, TA1), 3 (October 2015, TA2; September 2016, TA3), and 6 (November 2018, TA4) with 35 to 46 antennas of the 12\,m main array in one frequency setting in each of Band~6 (TA1--TA3; 232\,GHz) and Band~7 (TA4; 354\,GHz). In both settings, the data set contains four 1.875\,GHz spectral windows with 1920 channels each, except for one spectral window in the cycle~3 data that has 960 channels covering 1.875\,GHz. The final velocity resolutions are 3.0 and 1.0\,km\,s$^{-1}$ for the Band 6 and 7 data, respectively. The baselines range from 13 to 16196\,m for band~6 and 15 to 1398\,m for Band~7. This means a highest angular resolution of about 0\farcs03 and 0\farcs2 for the two bands, respectively. Bandpass calibration was performed on J1107-4449 (TA1, TA2, TA3) and J1037-2934 and J0635-7516 (TA4), and gain calibration on J1131-5818 (TA1), J1132-5606 (TA2, TA3, TA4). Flux calibration was done using Ceres and Titan (TA1), J1131-5818 (TA2), J1150-5416 (TA3), and J0635-7516 and J1037-2934 (TA4). Based on the calibrator fluxes, we estimate the absolute flux calibration to be accurate to within 5\%. 

The data were reduced using different versions of the Common Astronomy Software Applications package (CASA; \citet{mcmuetal07}). After corrections for the time and frequency dependence of the system temperatures, and rapid atmospheric variations at each antenna using water vapour radiometer data, bandpass and gain calibration were performed. For the Band~6 setting, data obtained in three different configurations were combined. Subsequently, self-calibration was performed on the strong continuum for the Band~6 data. Imaging was carried out using the CASA {\tt tclean} algorithm after a continuum subtraction was performed on the emission line data. The final line images were created using Briggs robust weighting (0.5). This resulted in beam sizes of 0\farcs028$\times$0\farcs026 (38$^\circ$) and 0\farcs22$\times$0\farcs17 (--48$^\circ$) for the Bands~6 and 7 data, respectively. Typical rms noises per channel are $\approx$\,0.6 and $\approx$\,1.3\,mJy\,beam$^{-1}$ for the Bands~6 and 7 data at 3 and 1\,km\,s$^{-1}$ resolution, respectively.

%
%
%
%
 
\subsection{Archival optical spectra}
\label{s:opt_obs_desc}

The archive of the European Southern Observatory (ESO) contains some excellent high-resolution spectra of HD\,101584. These spectra were obtained with the ESO VLT/UVES instrument through the observing program 266.D-5655(A) and were retrieved by us as pipeline-processed data products. All 19 exposures were made on 2003-02-04 between 09:00 and 09:30 UT (JD 2452674.9). The spectra in the visual and red region (wavelengths 4727--10426\,\AA) have a resolving power of $R$\,=\,74450, corresponding to a velocity resolution of 4.0\,km\,s$^{-1}$. The blue spectra (3760--4990\,\AA) have $R$\,=\,65030 or a resolution of 4.6\,km\,s$^{-1}$. We performed a detailed analysis of small wavelength intervals  to search for emission lines that might be associated with the region that emits the observed mm-wave emission lines.

%
%
%
%
\section{Line identification}

\begin{figure}
\centering
   \includegraphics[width=8cm]{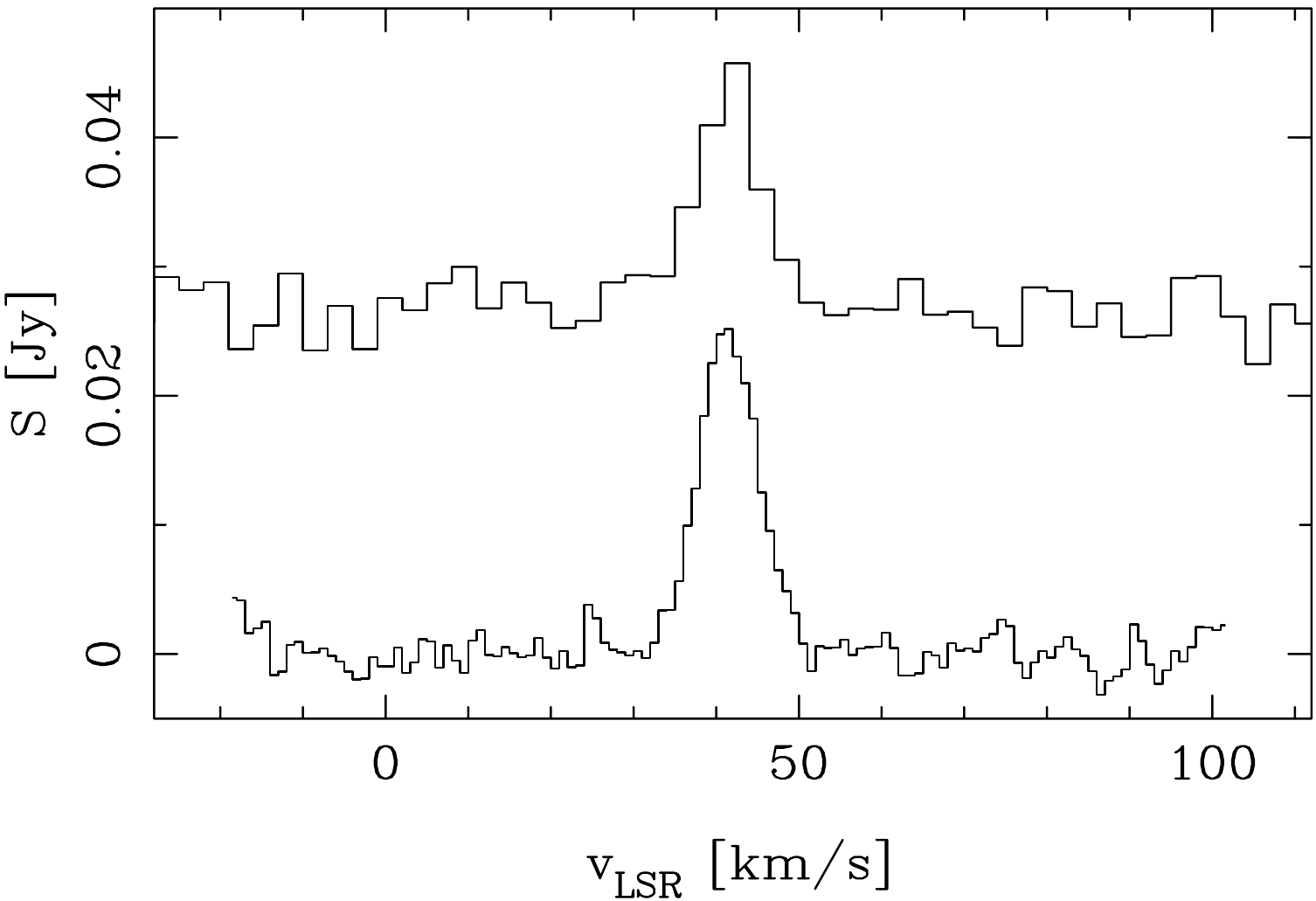}
    \caption{X30$\alpha$ (upper spectrum; scaled by a factor of three and biased by 25 mJy, 3\,km\,s$^{-1}$ resolution) and X26$\alpha$ (lower spectrum; 1\,km\,s$^{-1}$ resolution) lines observed towards HD\,101584. Continuum emissions have been subtracted. 
    }
   \label{f:lines}
\end{figure}

We detect two lines, at rest frequencies of 232.0245 and 353.8100\,GHz (adopting a systemic velocity of 41.5\,km\,s$^{-1}$) in our ALMA observations towards HD101584; see Fig~\ref{f:lines}. Both have Gaussian line shapes with full widths at half maximum (FWHMs) of about 9\,km\,s$^{-1}$. Both line emissions are spatially unresolved and centred on the continuum peak, which is believed to be the position of the star. The best upper limit to the line-emitting source size is obtained for the 232.0245\,GHz line, with a FWHM\,$\la$\,0\farcs02 assuming a Gaussian source (after deconvolving with the beam).  

The source is modestly rich in molecular lines with a line density of less than three lines per GHz \citep{olofetal19}, but we have found no identification in terms of relevant molecular lines.\footnote{There is an H$_2^{13}$CO(\mbox{$5_{05}-4_{04}$}) line at 353.8119\,GHz. However, this can be ruled out as an identification because we know from previous data that the H$_2$CO lines are strongest at the so-called extreme velocity spots \citep{olofetal17, olofetal19}, and the H$_2^{13}$CO(\mbox{$5_{05}-4_{04}$}) line is not detected in any of them.} Furthermore, the line widths  differ from those of the molecular lines obtained within an aperture of about 0\farcs2, Sect.~\ref{s:predictions}. On the other hand, these rest frequencies are consistent with an identification in terms of the X30$\alpha$ and X26$\alpha$ RTLs from an element X heavier than carbon. As an example, the S30$\alpha$ and S26$\alpha$ lines lie at 232.0233 and 353.8093\,GHz. In Sect.~\ref{s:model} we present a simple model that makes the identification in terms of atomic RTLs viable. Nearby X$n\beta$ lines lie outside the frequency coverage of our observations.

\begin{table*}[t]
\caption{Observed line and continuum characteristics.}
\begin{tabular}{l c c c c c c c}
\hline \hline
Line       & $\nu_{\rm rest}$\,$^a$ & $S_{\ell}$  & $\Delta \upsilon$\,$^b$ & $\int S_\ell {\rm d}\upsilon$ & $\theta_\ell$\,$^c$ & $S_{\rm c}$  & $\theta_{\rm c}$\,$^d$ \\
           & $[$GHz]          & [mJy]        & [km\,s$^{-1}$]    & [Jy\,km\,s$^{-1}$] & [\arcsec ] & [mJy]  &  [\arcsec] \\
\hline 
X30$\alpha$ & 232.0245\,$\pm$\,0.0016\phantom{0}  & \phantom{0}7.3$\pm$0.7 & 8.7$\pm$1.0   & 0.074\,$\pm$\,0.007 & $<$0.02            & \phantom{0}7.5$\pm$0.5     &  0.011$\pm$0.001  \\
H/He/C\,30$\alpha$ &         & $<$0.8 \\
X26$\alpha$ & 353.8100\,$\pm$\,0.00077  & 25.1$\pm$1.5           & 8.5$\pm$0.2   & 0.225\,$\pm$\,0.012 & $<$0.04            & 61.2$\pm$3.0     &  0.190$\pm$0.016\\
H/He/C\,26$\alpha$ &         & $<$1.2 \\
\hline
\end{tabular}
\label{t:obs_rec_res}
\tablefoot{$^{(a)}$ Assuming emission at the systemic velocity of 41.5\,$\pm$\,0.15\,km\,s$^{-1}$. $^{(b)}$ FWHM of a Gaussian fit to the line profile (deconvolved with the spectral resolution). $^{(c)}$ Upper limit to the source size (FWHM of a Gaussian source after deconvolving with the beam) using a Gaussian fit to the line brightness distribution in the systemic velocity channel. $^{(d)}$ Estimated source size (FWHM of a Gaussian source after deconvolving with the beam) using a Gaussian fit to the continuum brightness distribution. The 232\,GHz data are obtained with a circular beam of 0\farcs025, while those at 353\,GHz are obtained with a beam of 0\farcs24$\times$0\farcs17 (19$^\circ$) and are likely to have a significant contribution from dust continuum emission.}
\end{table*}

Direct identification of the carrier(s) is more difficult because the RTLs from heavy elements are close together in rest frequency, for instance, the $^{23}$Na and $^{56}$Fe 30$\alpha$ and 26$\alpha$ lines lie at 232.0217 and 232.0250\,GHz and 353.8069 and 353.8112\,GHz, respectively. That is, they lie on each side of our determined rest frequencies, and well within the observed line widths that correspond to $\approx$\,7 and 11\,MHz for the X30$\alpha$ and X26$\alpha$ lines, respectively. Therefore, theoretical modelling is required to make further progress. 

A useful formula for calculating rest frequencies of RTLs (ignoring the fine structure of hydrogen and the electronic structure of the core for non-hydrogenic atoms), $n+\Delta n$\,$\rightarrow$\,$n$, is given by the Rydberg formula of \citet{towletal96},
\begin{eqnarray}
\label{e:recfreq}
\nu_{n+\Delta n,n} & = & 3.289842\times10^6\, (1-5.48593\times10^{-4}/m_{\rm a}) \times \nonumber \\
                   &   &\times\,\left[\frac{1}{n^2} - \frac{1}{(n+\Delta n)^2} \right]\,\, [{\rm GHz}], 
\end{eqnarray}
where $m_{\rm a}$ is the atomic mass of the species in atomic mass units (in the usual nomenclature $\Delta n$\,=\,1 is an $\alpha$ line, $\Delta n$\,=\,2 a $\beta$ line, etc.).

The continuum flux from the line-emitting source is difficult to estimate due to contamination with dust continuum emission. The Band 6 data have the highest resolution and using different uv-tapering we estimate flux densities of 7.5, 10.1, and 13.2\,mJy for beam sizes of about 0\farcs025, 0\farcs083, and 0\farcs15, respectively. This gives a flux density that increases only slowly with beam area, which suggests that the 7.5\,mJy within an estimated source size of 0\farcs011 (the FWHM of a Gaussian source after deconvolving with the beam) is relatively free from dust emission at this frequency.\footnote{The contribution to the continuum from the visible photosphere of a star with the characteristics of HD\,101584 is negligible.} The Band 7 continuum source size is 0\farcs19, as measured by an 0\farcs23 beam, and the flux density is 61\,mJy. Even if the dust contribution increases only slowly with beam area, the contribution from dust continuum is likely to be significant in this case.

The line characteristics are summarised in Table~\ref{t:obs_rec_res}. The line strengths are 7.3 and 25\,mJy and the line-to-free-free-continuum flux density ratios are $>$\,1.0 and $>$\,0.4 for the 30$\alpha$ and 26$\alpha$ lines, respectively. Upper limits to the corresponding H, He, and C lines are 1 and 2\,mJy in the two bands, respectively. That is, for the 26$\alpha$ line the X/H line intensity ratio is $\ga$\,10. The lines are perfectly Gaussian, within the noise levels, and there is no evidence of low-intensity, broad line wings.

%
%
%
%
\section{A simple model in strict LTE}
\label{s:model}

The atomic lines reported here in the mm-wave spectrum of HD\,101584 arise in the immediate environment of the binary system, within a few astronomical units, as shown by the observations. A complete analysis of the data requires a non-LTE model that takes into account chemistry, ionisation, excitation, and radiative transfer in a source of arbitrary geometry. However, this is far beyond the scope of this paper. Instead, we have devised a model in strict LTE that will explain the presence of heavy-element RTLs and a free-free continuum, at their observed intensities, and the simultaneous absence of the corresponding C and H RTLs. With such a model it is possible to construct models of the RTL emission, from an assumed geometry of the gas, with the smallest number of free parameters. The model can also be used to easily make predictions for future observations. Because of the complexity of the problem, it is difficult to argue in detail whether or not such an approach is justified, but  here we present some arguments in favour of the LTE model.

In our opinion, the primary star with its modest luminosity ($\approx$\,1600\,[$D$/1\,kpc]$^2$\,$L_\odot$) and moderate temperature ($\approx$\,8500\,K) is not able to produce an ionised nebula that can explain the observed RTL and continuum flux densities and, at the same time, the absence of the corresponding C and H RTLs. First, the high density that is required to explain the observed line and continuum flux densities (see below) will lead to efficient absorption of the ionising radiation already close to the star (the density is much higher here than in the traditional nebular case). The alternative, that the RTL region is located at a distance from the star with an absorption-free cavity in between, appears less likely. Second, such a nebula, produced by a star with the characteristics of HD\,101584, would produce H and C RTLs that are stronger than the corresponding heavy-element lines, in stark contrast to the observed result. 

The question remains as to whether or not strict LTE is well justified. In order for the ionisation, chemical abundances, internal excitation, continuum brightness, and line intensities to be well characterised by equilibrium conditions, the rates of all microscopic processes must be in detailed balance at the same value of the temperature. The densities of the best-fitting LTE models (see below), $n_{\rm H}$\,$\approx$\,10$^{11}$ --\,10$^{12}$\,cm$^{-3}$, are higher by orders of magnitude than typical densities in the interstellar medium where non-equilibrium prevails. On the other hand, this density is still much lower than the density in a typical stellar photosphere at the same temperature, where LTE is a useful starting approximation, but where subtle non-LTE effects can nevertheless be found in ionisation and line formation of some elements. Simple estimates of chemical, photodissociation, and ionisation timescales, as well as comparisons of collisional and radiative strengths, suggest that the LTE assumption is, if not perfect, at least acceptable. In addition, the line flux density ratio, 26$\alpha$/30$\alpha$\,=\,3.4$\pm$0.4, is not too far from the blackbody ratio (in the Rayleigh-Jeans limit) for emission at 232 and 354\,GHz, namely 2.3, and we therefore suspect that the lines are formed in thermal equilibrium. The slightly higher value for the observed flux density ratio than the blackbody ratio can be attributed to an optical depth effect, that is, the lines are at least partially optically thin and the optical depth of the 26$\alpha$ transition is higher than that of the 30$\alpha$ transition. 

Consequently, we assume that the abundances of molecular species, the ionisation balance, and the level populations are all governed by the processes that tend to drive the gas toward the equilibrium state at the kinetic temperature prevailing in the near circumstellar environment, that is, we assume that strict local LTE applies. This assumption is also the reason behind our choice to  deliberately refer to the spectroscopic features as RTLs rather than radio recombination lines  throughout this
paper in order to steer the reader away from any prejudice that the lines arise mainly through the process of radiative recombination or that they should behave like the well-known features in nebular spectra. Further, the equilibrium temperature must be low enough that ionisation of hydrogen, helium, and carbon does not occur. We compare the observational constraints (Table~\ref{t:obs_rec_res}\,\footnote{The observations of the two lines were made at different times; therefore, it is possible that variability has affected the ratio of their fluxes. Even so, we treat the measurements as though they were contemporaneous and assume that both lines arise in the same emitting gas.}) to a simple model of a uniform, isothermal parcel of gas in strict LTE. This parcel of gas is characterised by two parameters: a number density of hydrogen in all forms, $n_{\rm H}$, and a temperature, $T$. To this should be added the parameters defining the geometry and kinematics. Once the true geometry and kinematics of the source have been established, a more sophisticated modelling can be attempted.

%
%
%
%
\subsection{Geometry}

We have chosen a geometry in light of the VLTI/PIONIER results of \citet{klusetal20}, who infer a face-on (as seen from Earth), circumbinary dust disc  with a sublimation rim at about 2\,[$D$/1\,kpc]\,au from the central object. The region in which the RTLs are formed could therefore consist of dust-free gas surrounding the binary system and extending along the upper and lower surface layer of this dust disc (the estimated source size is larger than the diameter of the dust rim, see below). 

Considering this, we have chosen a simple geometry in terms of a cylinder seen along its axis, where $R$ is the radius of the source in the plane of the sky and $\Delta z$ is the depth of the cylinder along the line of sight. The apparent angular diameter of the emitting region is $\theta$\,=\,2$R/D$, and we specify the geometry of the emitting region by the parameter $\rho$\,=\,$\Delta z/R$. The case $\rho$\,=\,1 corresponds to a cylinder of radius $R$ and height $\Delta z$\,=\,$R$ viewed face-on at a distance $D$. When $\rho$\,$\ll$\,1, the modelled flux refers to that of a very thin disc, while the $\rho$\,=\,2 case is close to that of a spheroid. We adopt $\rho$\,=\,0.1, that is, a geometrically thin disc, as the most likely geometry for describing the RTL region. This also minimises the volume of the relatively warm gas required to explain the observations.

In strict LTE, it is possible to derive a lower limit to the size of the line-emitting region. The flux densities of the lines cannot exceed that of a blackbody at a given temperature. For a temperature of 3000\,K (it is shown below that this is a reasonable temperature in this context), both lines give about 10\,mas as the lower limit to the source size. This agrees very well with the measured Gaussian size of the continuum source at 232\,GHz, namely 11\,mas. This is our best estimate for the source size because the 232\,GHz data have a much higher angular resolution than those at 354\,GHz, and the continuum data have a much higher S/N ($\approx$\,70) than the line data ($\approx$\,13). We make the reasonable assumption that the line and continuum emissions come from the same region. As shown by \citet{vanho00} the relation between the true source size and the size of a Gaussian fit to its brightness distribution depends on the geometry of the source; for instance, a uniform brightness disc is about 1.5 times larger than the FWHM of the Gaussian fit to its brightness distribution. We will nevertheless adopt 11\,mas as the size of the RTL region, that is, the assumed thin disc extends to $R$\,=\,5.5\,[$D$/1\,kpc]\,au. The reason is twofold: first, and most importantly, the resulting density and temperature depend only weakly on the source size (see Sect.~\ref{s:geom_dist} and Appendix~\ref{a:geom_dist}), and secondly, in our case the true source geometry is unknown.

Of course, other geometries are not excluded, keeping in mind that this is at the heart of the complex circumstellar morphology as shown by our ALMA data \citep{olofetal19}. In all likelihood, an accretion disc, an associated jet, and the innermost part of an hourglass structure are located in the region \citep{olofetal19}. However, as shown in Appendix~\ref{a:geom_dist}, results such as mass, density, and temperature of the gas depend little on the assumed geometry.

%
%
%
\subsection{Chemistry}

We consider an ideal gas composed of hydrogen and the 14 other most abundant elements of comparable or smaller first ionisation potential. Helium is included, although it remains fully neutral at the temperatures of interest here. Two different sets of abundances are compared in the analysis, as summarised in Table~\ref{t:abundances}. The reference abundances are those of the solar photosphere ($Y$; Set~1), and Set~2 consists of abundances depleted by a factor of $d$. 

In addition to the elemental composition we take into account the formation of some simpler molecules, as this may have important consequences for the atomic abundances, the atomic ionisation balance, and for the dominant sources of continuous opacity in the gas, which depend on collisions between free electrons and ions and the principal neutral species H, H$_2$, and He. We therefore calculated the molecular abundances (at each grid point) for the species given in Table~\ref{t:chemistry}\,\footnote{In the chemistry calculations, the elements N and O also play important roles. They are assumed to take on their solar abundances.}. We used the most recent values for partition functions and dissociation energies in the calculation of the equilibrium abundances. The  molecular partition function is a sum over all low-lying electronic states and enough of their vibrational and rotational levels to yield a converged value at temperatures of interest. The full list of included molecules, their dissociation energies, and abundances in the reference model is presented in Table~\ref{t:chemistry} of Appendix~\ref{a:chemistry}.

It should be noted here that recombination lines from molecules can be ignored because the recombination of a molecular ion and an electron preferentially leads to a dissociation of the molecule. 

%
%
%
\subsection{Ionisation balance}

All elements X$_i$ are taken to exist in two ionisation states, the neutral atom X$_i^0$ and first ion X${_i^{+1}}$, except hydrogen, which also forms a potentially important anion, H$^-$. Each internal energy state in the gas is labelled with the indices $i$ for the element, $j$\,=\,0,$\pm$1 for the charge state, and $k$ for the quantum state of energy $E_{ijk}$. In strict LTE, the population of state $ijk$, $p_{ijk}$, can be expressed as a number density given by the Boltzmann formula,
\begin{equation}
\label{e:eq1}
\frac{p_{ijk}}{p_{ij}} = \frac{g_{ijk} \exp (- E_{ijk}/k_{\rm B} T)}{Q_{ij}(T)}\,,  
\end{equation}
where
\begin{equation}
Q_{ij}(T) = \sum_k g_{ijk} \exp (- E_{ijk}/k_{\rm B} T)  
\end{equation}
is the partition function and $p_{ij}$ the total density of the ion X$_i^j$ summed over all its states, $p_{ij}$\,=\,$\sum_k p_{ijk}$. For each element $i$, the sum over all states of all its ions is related to the total number density of hydrogen in all forms, $n_{\rm H}$, by
\begin{equation}
\sum_j p_{ij} \equiv \sum_j \sum_k p_{ijk} = n_{\rm H} 10^{(Y_i - Y_{\rm H})}  \,\, .
\end{equation}

The ionisation balance is governed by the Saha equation
\begin{equation}
 \frac{p_{i1} p_e}{p_{i0}} = \frac{Q_{i1} Q_e}{Q_{i0}} \exp (-I_i/k_{\rm B} T ) \,\,,
\end{equation} 
where
\begin{equation}
Q_e = \left(\frac{2\pi m_e k_{\rm B} T}{h^2}\right)^{3/2} 
\end{equation}
is the partition function of the continuum of free-electron states (translational energies), $I_i$ is the first ionisation potential of element $i$, $k_{\rm B}$ is Boltzmann's constant, $h$ is Planck's constant, and $m_{\rm e}$ is the electron mass. Overall charge neutrality is assumed, so that the density of free electrons is given by
\begin{equation}
\label{e:eq6}
p_e = \sum_i p_{i1} - p_{{\rm H},-1}  \,\, . 
\end{equation}
Given a set of element abundances $Y_i$ and the two parameters $T$ and $n_{\rm H}$, Eqs. (\ref{e:eq1}) to (\ref{e:eq6}) are solved simultaneously by iteration. The result is a complete set of densities, $p_{ijk}$. We computed the partition functions of neutral atoms and first ions by direct summation over tables of energy levels derived from the {\it NIST Atomic Spectra Database}\,\footnote{The U.S. National Institute of Standards and Technology (NIST) maintains a critically reviewed database of atomic energy levels at {\tt https://physics.nist.gov/PhysRefData/ASD/ \\levels{\_}form.html} }. 

%
%
%
\subsection{Radiative transfer}

Once the populations have been determined, the continuum opacity can be evaluated. At radio, mm, submm, and far-infrared wavelengths, the important processes in the weakly ionised gas are electron--ion bremsstrahlung (free-free), electron--neutral bremsstrahlung, and Thomson scattering. The electron--ion free-free absorption coefficients are computed using the accurate Gaunt factors of \citet{vanhoetal14}. We compute the electron--neutral (e$^-$\,+\,H) free-free absorption coefficients based on the analytical approximation of \citet{stal74}, which agrees within 5\% with accurate calculations \citep{bellberr87, john94b} at wavelengths $\lambda$\,$>$\,1\,$\mu$m and temperatures $T$\,=\,1400 to 10080\,K. We also include $e^-$\,+\,He \citep{john94a, john94d} and $e^-$\,+\,H$_2$ \citep{john94c} in the  electron--neutral free-free opacity. In principle, electrons on polar, neutral molecules like CO and H$_2$O produce very large free-free opacities per molecule \citep{john75a, john75b}; however, they contribute relatively little to the total compared with the more abundant H, H$_2$, and He in our models. In order to describe the opacity at mid-infrared and shorter wavelengths ($\lambda$\,$<$\,50\,$\mu$m), we include non-relativistic Thomson scattering by free electrons and H$^-$ bound-free absorption (cross-sections from \citet{mclaetal17a, mclaetal17b}).

These absorption coefficients are conventionally expressed as an inverse length per unit density of the atom or ion and per unit electron pressure, $\kappa$, where the electron pressure is $P_{\rm e}$\,=\,$p_{\rm e} k_{\rm B} T$. The optical depth in the continuum through a uniform medium of length $\Delta z$ is thus
\begin{equation}
\tau_{\rm c} = P_{\rm e}\, ( p_{\rm H} \kappa_{\rm e,neutral} + p_{\rm e} \kappa_{\rm e,ion} )\,  \Delta z  \,\, . 
\end{equation}
Under the assumption of a Gaussian line profile with FWHM $\Delta \upsilon$, the corresponding optical depth at the centre of a line of element $i$ in ion stage $j$ is given by
\begin{equation}
\tau_{n'n''} = 3.738\times 10^{-7} A_{n'n''} p_{ijn''} \left( \frac{g_{n'}[1 - \exp(-h\nu /k_{\rm B} T)]}{g_{n''} (\nu /c)^3 \Delta \upsilon} \right)  \Delta z \,\, , 
\end{equation} 
where $g_{n'}$ and $g_{n''}$ are the statistical weights of the respective upper and lower states of the transition $n' \to n''$, $A_{n'n''}$ is the spontaneous transition probability [s$^{-1}$], $c$ is the speed of light [cm\,s$^{-1}$], and the line width has dimensions [km\,s$^{-1}$]. The product $p_{ijn''} \Delta z$ is the lower-state column density with dimensions [cm$^{-2}$].  With the assumption of LTE, the excitation temperature of any transition is equal to $T$, the source function is equal to the Planck function $B_{\nu}(T)$, and the observable flux density of either the line peak or continuum is
\begin{equation}
\label{e:eq9}
S_{\nu} = B_{\nu}(T) [1 - \exp(-\tau)]\, \pi \left( \frac{R}{D} \right)^2  \,\, ,
\end{equation}
that is, the line and continuum emissions are treated separately. This requires, for instance, optically thin emission, a situation that applies reasonably well to our models. To summarise the model, Eqs. (\ref{e:eq1}) through (\ref{e:eq9}) show how the observable flux densities of line and continuum depend on the parameters $n_{\rm H}$, $T$, $\theta$ (or $R$), $\rho$, and $D$ for a specified set of elemental abundances, where $n_{\rm H}$\,=\,$n_{{\rm H}^0} + n_{\rm H^+} + n_{\rm H^-} + 2(n_{\rm H_2}+n_{\rm H_2^+}) + 3n_{\rm H_3^+}$.

%
%
%
\subsection{Line broadening: Doppler motions and blending of emitters}

The shapes and widths of the line profiles contain valuable information. The observed lines are fit very well by single Gaussian functions, with a FWHM of $\Delta \upsilon$\,$\approx$\,8.5\,km\,s$^{-1}$ (taken from the X26$\alpha$ line that is observed with the highest spectral resolution). It is significant that the $26\alpha$ and $30\alpha$ lines show the same value of line width, and that this line width is substantially larger than the thermal line widths at the relevant temperatures (e.g. for neutral Mg it is only 2.4\,km\,s$^{-1}$ at $T$\,=\,3000\,K). 

Typically this would be explained by a combination of thermal Doppler broadening and micro-turbulence, with only minor contributions from any larger scale kinematical motions. For simplicity, we adopt this description in the determination of the best-fit equilibrium model, the reference model, and separate the two components into thermal Doppler motions ($\Delta \upsilon_{\rm th}$) and micro-turbulent motions ($\Delta \upsilon_{\rm turb}$). We further assume that both have Gaussian broadening functions along the line of sight through the emitting region. The Gaussian widths add in quadrature $\Delta \upsilon^2$\,=\,$\Delta \upsilon_{\rm th}^2$ + $\Delta \upsilon_{\rm turb}^2$.

Broadening of the lines results also from the fact that each line is most likely a blend of lines due to a mixture of different species. The slight differences in reduced mass of this species lead to slight differences in line frequencies; see Eq.~(\ref{e:recfreq}). These velocity differences are independent of $n$ for any $\alpha$ transition.

In the modelling, we obtained solutions by linear combination of optical depths of overlapping lines of the different atoms in the Rydberg approximation. The best-fitting models have line optical depths that are less than or of the order of unity, which justifies simply summing the optical depth at each point on a densely sampled frequency grid. In the construction of a theoretical profile, each of the line components is weighted by the relative abundance of the atomic species.

Rydberg atoms are very sensitive to perturbations by collisions with other atoms and electrons and by magnetic forces. Line profiles broadened by such perturbations provide limits on gas density and magnetic field strength as discussed in Appendix~\ref{a:add_broadening} and Sect.~\ref{s:line_broadening} below.

%
%
%
%
%
\section{Results from modelling mm-wave RTLs}

The model described in the preceding section can be used to gain insight into the conditions that make mm-wave RTLs observable towards stars that are only moderately warm. We have several observable properties and a model with four free parameters. We start by explaining the adopted procedure for finding the best-fit model.

\subsection{Procedure for selecting the best-fit model}

To find the model that best fits the observations is not trivial, and therefore we describe the procedure employed:

\begin{itemize}
\item{Adopt a distance $D$ and a set of elemental abundances.}
\item{Select a value of the geometrical parameter $\rho$.}
\item{For one line, specify the peak flux density $S_{\nu}$ and observed line width $\Delta \upsilon$.}
\item{Specify a grid of values of the parameters $n_{\rm H}$ and $T$.}
\item{For each set of density and temperature, compute the peak line flux densities for a list of lines, recalling that the observed line might be an unresolved blend of several emitters.}
\item{Iterate over values of the source radius $R$ until the model matches the line flux.}
\item{Examine the resulting grid of solutions to see which of them also satisfies the upper limit on the hydrogen line flux, the flux of the adjacent continuum, and the observed constraints on angular size of the source.}
\item{For the same choice of $D$, abundances, and $\rho$, repeat the above steps for a second transition.}
\item{Determine whether there is some parameter space that satisfies both transitions and their continuum fluxes simultaneously.}
\item{Repeat the entire procedure for different choices of elemental abundances and $\rho$.}
\end{itemize}

The line list for the RTLs includes two lines (H and C) and one blend (all abundant heavier elements that contribute to the blend, Na to Ni). The densities of the neutral species have been computed explicitly in strict LTE. The mathematical model is non-linear in several regards. The temperature enters in the arguments of exponential functions in the Saha and Boltzmann equations describing the ionisation balance and excitation. The source path length $\Delta z$\,=\,$\rho R$ is a term in each optical depth and thus enters in the arguments of exponential functions. The coupled parameter for the radius enters as $R^2$ in the flux densities. Thus, the solutions must be found by successive iterations. Remarkably, in the present case, the solutions appear to converge on a relatively small volume in parameter space.

%
\begin{figure*}
\centering
\includegraphics[width=8.0cm]{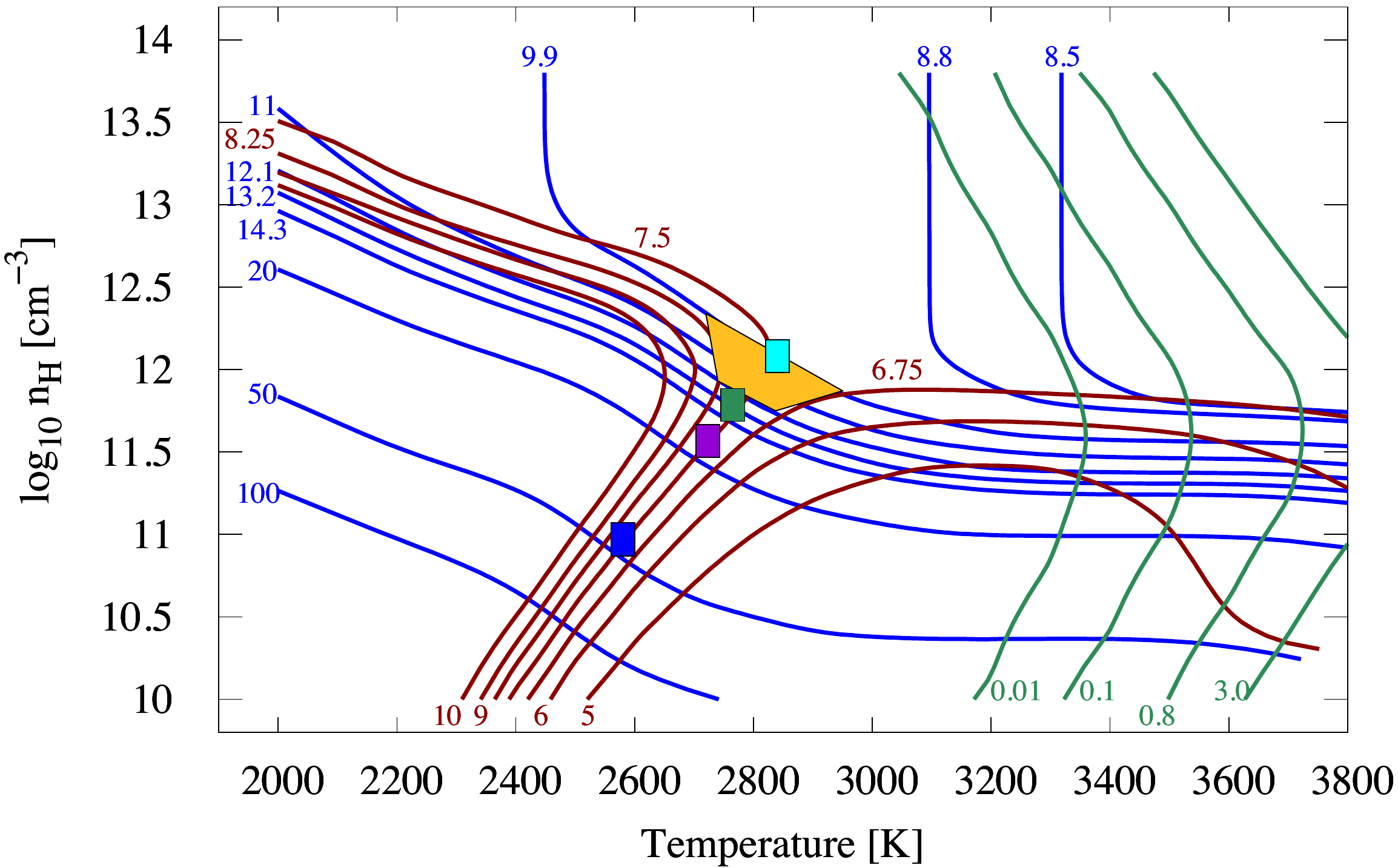} \hspace{2mm} \hspace{5mm} \includegraphics[width=8.0cm]{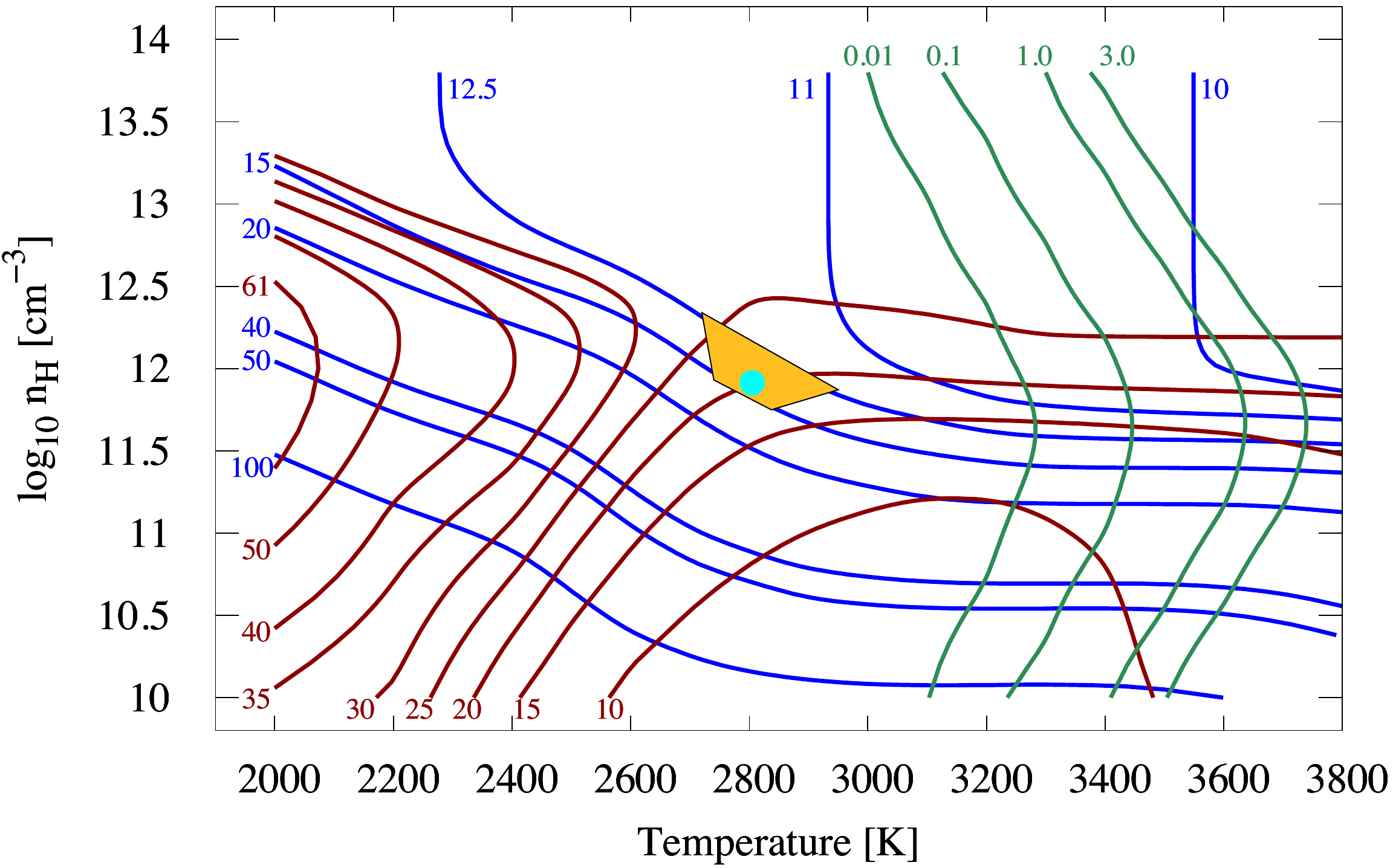}
\caption{{\bf Left:} LTE solutions with $\rho$\,=\,0.1 that yield a line flux density of 7.3\,mJy in the  $30\alpha$ blend for solar abundances. Blue contours display equal values of the angular diameter $\theta$ in mas. Dark red contours show the 232\,GHz continuum flux density in mJy. Green contours show the flux density of the corresponding hydrogen line in mJy. The polygon covers the space of primary solutions that allow for $\pm$\,10\,\% changes on the continuum flux density and the angular diameter for illustrative purposes. The shaded rectangles indicate how the primary solution shifts when the value of $\rho$ is changed by a factor of two (0.05 (violet) and 0.20 (green)), and when a source size of 17\,mas and $\rho$\,=\,2 are adopted (blue). {\bf Right:} Same as left panel but for the $26\alpha$ blend and a line flux density of 25\,mJy, and the 354\,GHz continuum flux density in mJy. The cyan solid circle indicates the location of the reference model.} 
\label{f:paraplot}
\end{figure*}

\subsection{The reference model}

The behaviour of the models is illustrated in two contour diagrams shown in Fig.~\ref{f:paraplot}. These diagrams of the ($\log n_{\rm H}$,$T$)-plane display contours of equal angular diameter ($\theta$ in mas, blue), continuum flux density ($f_{\rm c}$ in mJy, brown), and hydrogen line flux ($f_{\rm H}$ in mJy, green). These models were computed for $\rho$\,=\,0.1 (a cylinder with a depth equal to a tenth of its radius), undepleted solar abundances, a nominal distance of $D$\,=\,1.0\,kpc, and the line flux densities given in Table~\ref{t:obs_rec_res}, that is, all points in the diagrams give the correct observed line intensities. However, the observational results will limit the area of acceptable solutions, for instance, the upper limits on the H$30\alpha$ and H$26\alpha$ lines imply that the temperature must be $T$\,$\la$\,4000\,K, for the adopted solar abundances, and the measured source size is about 11\,mas. 

The X30$\alpha$ transition provides the most restrictive constraints, because of the higher angular resolution of the observations. As shown in Fig.~\ref{f:paraplot}:\,left, there are two regions of the ($n_{\rm H}, T$) plane that satisfy all of the observed constraints of the $30\alpha$ transition and adjacent continuum. We pick the midpoint of the polygonic area in the left panel of  Fig.~\ref{f:paraplot} as the reference model. The physical characteristics of this region are $T$\,=\,2800\,K and $n_{\rm H}$\,=\,8$\times$10$^{11}$\,cm$^{-3}$, and its size is $R$\,=\,5.5\,au. The resulting gas mass is $\approx$\,7$\times$10$^{-4}$\,$M_\odot$, that is, only about 2\,\% of the estimated mass of the CCS component. The ionisation fraction is low, $\approx$\,3$\times$10$^{-5}$, and CO dominates over H$_2$ with [CO]/[H$_2$]\,$\approx$\,6 and [CO]/[H]\,$\approx$\,\,3$\times$10$^{-4}$ (a significant fraction of the C is bound in CO). The corresponding H and C recombination lines are more than a million times weaker than the heavy-element blend. The properties of the reference model are given in Tables~\ref{t:mod_results} and \ref{t:mod_results_ion_chem}. The corresponding solutions for the $26\alpha$ transition are displayed in the right panel of Fig.~\ref{f:paraplot}. Remarkably, the solutions for this transition cluster in nearly the same part of parameter space as for the $30\alpha$ transition.

The contours of the 232\,GHz continuum flux density do not close in the upper part of the diagram. As a result, there is a secondary solution at higher density (2.2$\times$10$^{13}$\,cm$^{-3}$) and lower temperature (2050\,K). However, this solution is implausible for two reasons. A very large continuum optical depth is required in order for this model to produce the observed line flux within the observed angular size. This would imply that the $\tau_{\rm c}$\,$\sim$\,1 surface of the continuum source is much more extended than the line emission, in contradiction to the 232\,GHz line and continuum maps. Secondly, the secondary solution implies much higher molecular abundances than the primary model because it is both denser and colder. The result is that the predicted LTE emission line fluxes of lines of CO and SiO in the observed bands are much larger than those observed; see Sect.~\ref{s:predictions}.

There are no reasonable solutions for the X30$\alpha$ line with depleted interstellar abundances. Depleted abundances of the heavier elements (especially Mg, Si, and Fe) cause the corresponding hydrogen line to be stronger, relative to the lines of the heavier elements, than in the undepleted case. The fluxes of H30$\alpha$ and/or the 232\,GHz continuum are too high over the entire parameter space explored ($T$\,=\,1500 to 10000\,K).

\begin{table}
 \caption{Properties of the reference model defined by $T$\,=\,2805\,K, $n_{\rm H}$\,=\,8.2$\times$10$^{11}$\,cm$^{-3}$, $R$\,=\,5.5\,au, and $M$\,=\,7.2$\times$10$^{-4}$\,$M_\odot$ ($\rho$\,=\,0.1).}
\begin{tabular}{ll}
\hline \hline
Flux densities        &   [mJy] \\
\hline 
$S_{\nu}$(X30$\alpha$)     &    \phantom{0}7.3   \\
$S_{\nu}$(H30$\alpha$)     &    \phantom{0}2.2$\times$10$^{-6}$  \\
$S_{\nu}$(C30$\alpha$)     &    \phantom{0}2.0$\times$10$^{-8}$  \\
$S_{\rm c}$(232\,GHz)      &   \phantom{0}7.5  \\
$S_{\nu}$(X26$\alpha$)     &   25\phantom{.000}    \\
$S_{\nu}$(H26$\alpha$)     &    \phantom{0}7.0$\times$10$^{-6}$  \\
$S_{\nu}$(C26$\alpha$)     &    \phantom{0}6.5$\times$10$^{-8}$   \\
$S_{\rm c}$(354\,GHz)      &   20\phantom{.000}  \\
\hline \hline
Optical depths \\
\hline
$\tau_{\ell}$(232\,GHz)         &    1.2 \\
$\tau_{\rm c}$(232\,GHz)        &    1.3 \\
$\tau_{\ell}$(354\,GHz)         &    1.1 \\
$\tau_{\rm c}$(354\,GHz)        &    0.7 \\
\hline \hline
Contribution to the X26$\alpha$ flux density \\
\hline
Element             &   [\%] \\
\hline
Na                  &   \phantom{0}6.6 \\
Mg                  &    74 \\
Al                  &    11 \\
Si                  &   \phantom{0}1.9  \\
Ca                  &   \phantom{0}8.3  \\
Fe                  &   \phantom{0}1.9  \\
\hline
\end{tabular}
\label{t:mod_results}
\end{table}

\begin{table}
 \caption{Ionisation balance and chemical abundances (by number) in the reference model defined by $T$\,=\,2805\,K, $n_{\rm H}$\,=\,8.2$\times$10$^{11}$\,cm$^{-3}$, $R$\,=\,5.5\,au, and $M$\,=\,7.2$\times$10$^{-4}$\,$M_\odot$ ($\rho$\,=\,0.1).}
\begin{tabular}{llll}
\hline \hline
Element  &      Neutral             &   First Ion              &  Molecule (leading) \\
\hline
 H       &         1.0              &  4.8$\times$10$^{-12}$   &  H$_2$\,\,\,\,\, 4.6$\times$10$^{-5}$ \\
 He      &   8.3$\times$10$^{-2}$   &  2.8$\times$10$^{-32}$ \\
 C       &   4.4$\times$10$^{-7}$   &  4.4$\times$10$^{-14}$   &  CO\,\,\, 2.7$\times$10$^{-4}$ \\
 N       &   2.2$\times$10$^{-5}$   &  9.3$\times$10$^{-18}$   &  N$_2$\,\,\,\,\, 2.3$\times$10$^{-5}$ \\
 O       &   2.2$\times$10$^{-4}$   &  8.9$\times$10$^{-16}$   &  CO\,\,\, 2.7$\times$10$^{-4}$ \\
 Na      &   2.3$\times$10$^{-10}$  &  1.7$\times$10$^{-6}$ \\   
 Mg      &   2.0$\times$10$^{-5}$   &  1.9$\times$10$^{-5}$ \\
 Al      &   3.7$\times$10$^{-8}$   &  2.8$\times$10$^{-6}$ \\
 Si      &   3.1$\times$10$^{-5}$   &  1.2$\times$10$^{-6}$    &  SiO\,\, 3.4$\times$10$^{-7}$ \\
 P       &   2.6$\times$10$^{-7}$   &  2.2$\times$10$^{-12}$ \\
 S       &   1.3$\times$10$^{-5}$   &  2.0$\times$10$^{-14}$ \\
 Cl      &   3.2$\times$10$^{-7}$   &  6.2$\times$10$^{-17}$ \\
 K       &   5.2$\times$10$^{-13}$  &  1.1$\times$10$^{-7}$ \\
 Ca      &   4.1$\times$10$^{-9}$   &  2.2$\times$10$^{-6}$ \\
 Ti      &   2.9$\times$10$^{-9}$   &  8.6$\times$10$^{-8}$ \\
 Fe      &   2.5$\times$10$^{-5}$   &  6.4$\times$10$^{-6}$ \\
 Ni      &   1.6$\times$10$^{-6}$   &  3.7$\times$10$^{-8}$ \\
 \hline
 e$^-$   &   3.4$\times$10$^{-5}$ \\
 H$^-$   &   4.4$\times$10$^{-13}$ \\
\hline
\end{tabular}
\label{t:mod_results_ion_chem}
\end{table}

\begin{figure*}
\centering
\includegraphics[width=8cm]{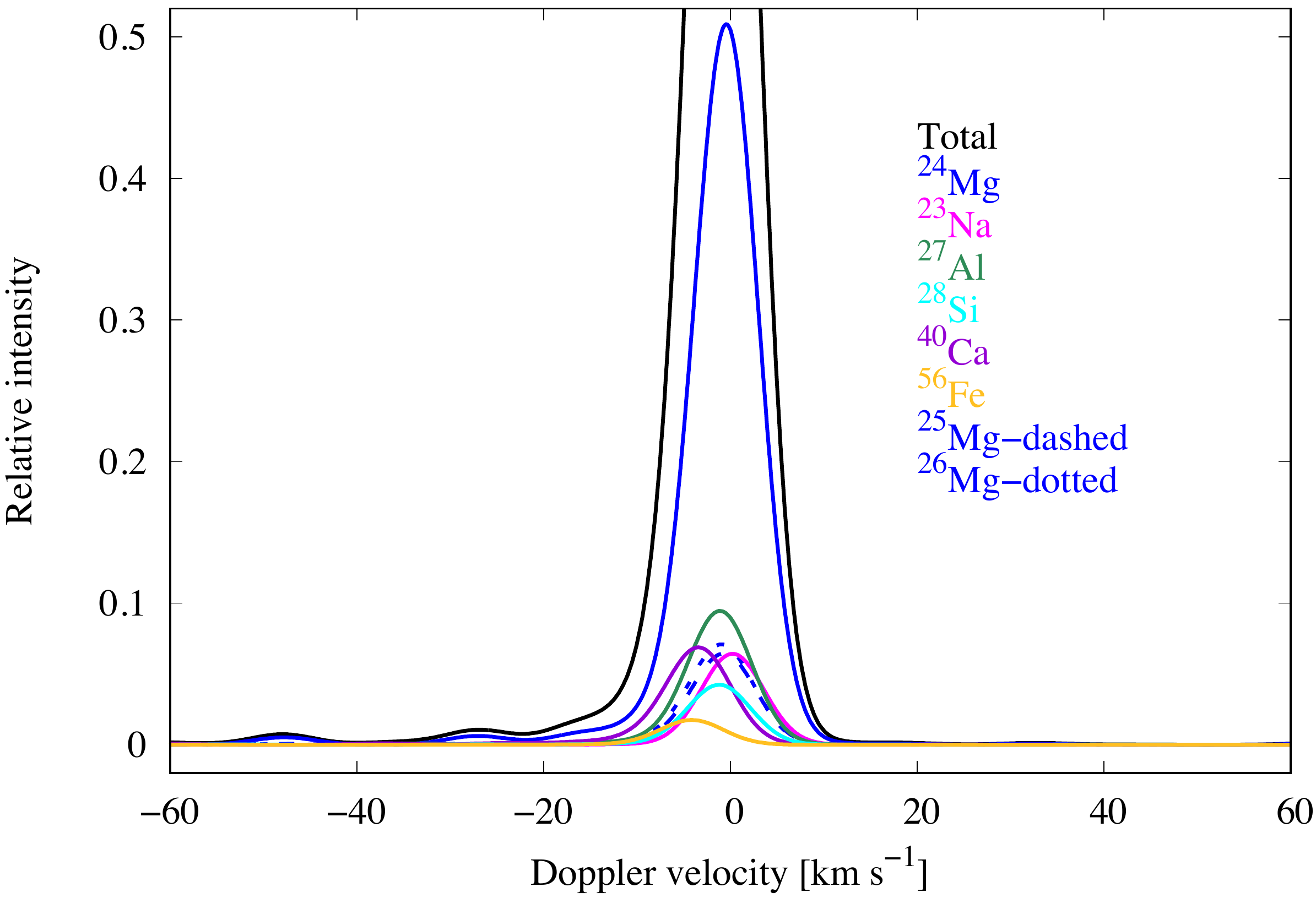} \hspace{1cm}  \includegraphics[width=8cm]{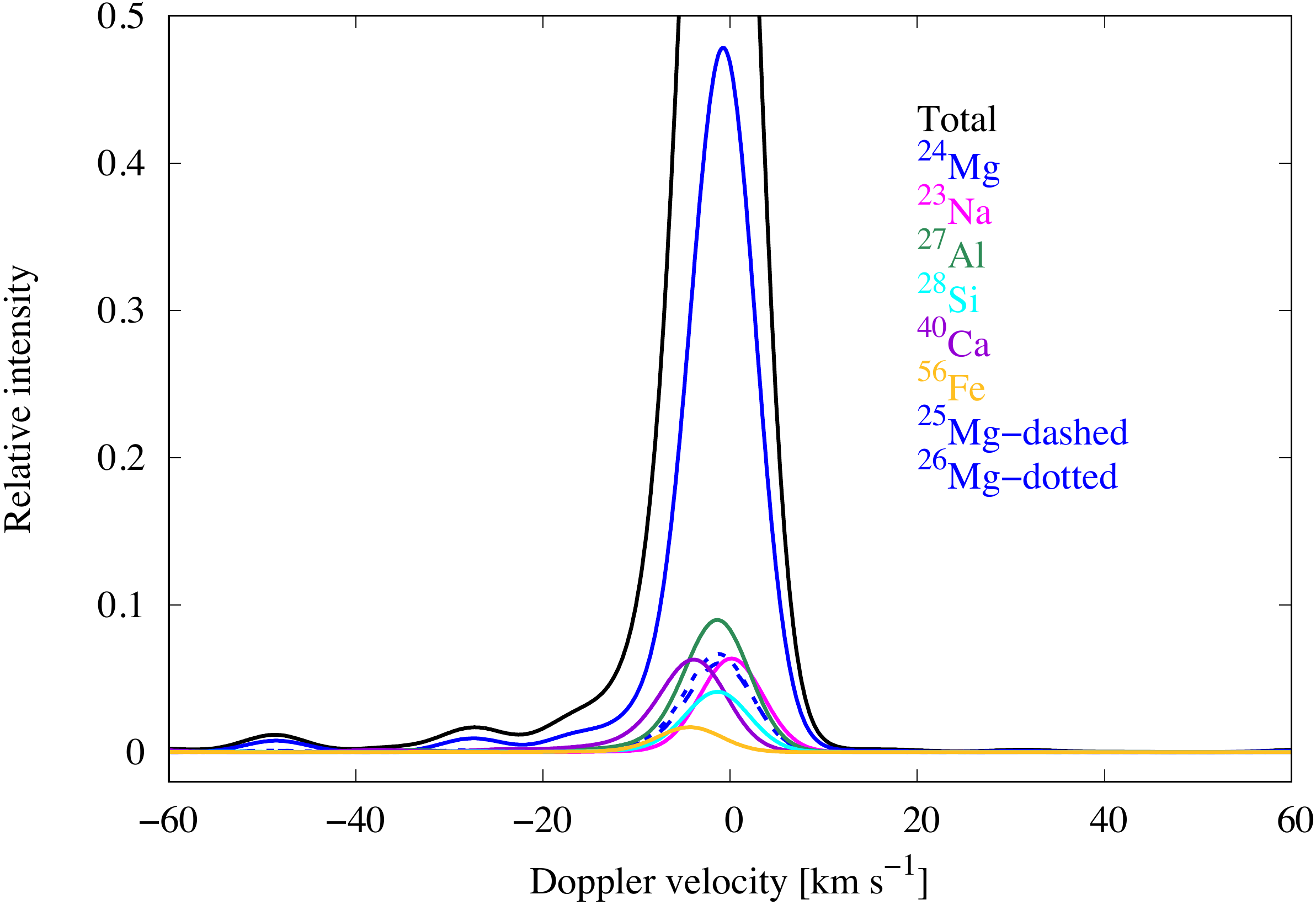}
\caption{Computed model profiles of the 30$\alpha$ (left) and 26$\alpha$ (right) transitions, excluding collisional broadening and radiative damping, are shown in black (truncated at a value of 0.5, the peak value is 1.0). The blended profiles of the eight atoms and isotopes are shown in colours to illustrate their relative contributions and the mass-shifts in frequency. All of the profiles are slightly asymmetrical owing to the fine structure as described in Appendix~\ref{a:add_broadening}. A turbulent velocity $\Delta \upsilon_{\rm turb}$\,=\,7.5\,km\,s$^{-1}$ is adopted so that the fitted FWHMs of the model profiles match the observed width, 8.5\,km\,s$^{-1}$.} 
\label{f:line_blend}
\end{figure*}

\begin{figure*}
\centering
\includegraphics[width=8cm]{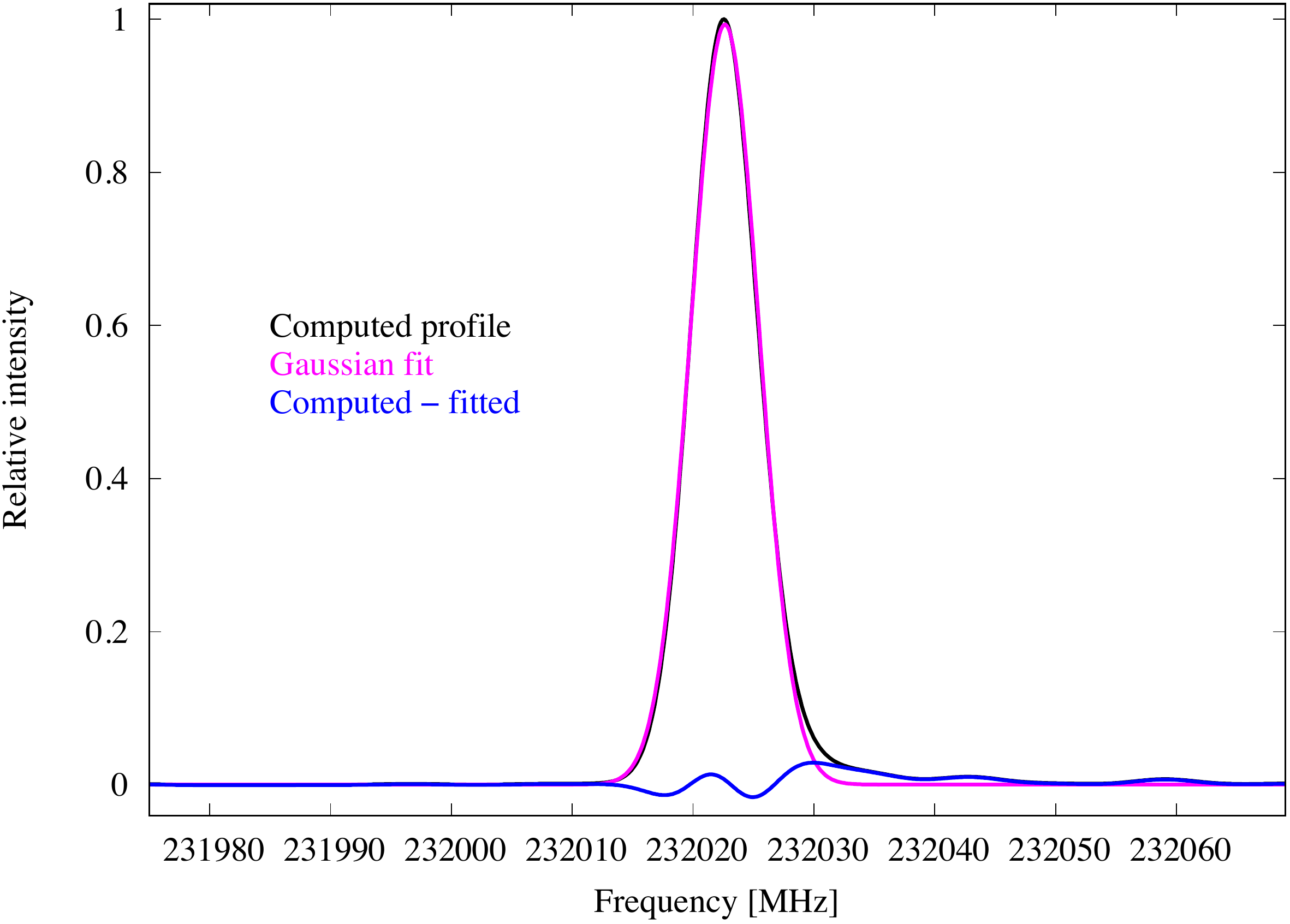} \hspace{1cm} \includegraphics[width=8cm]{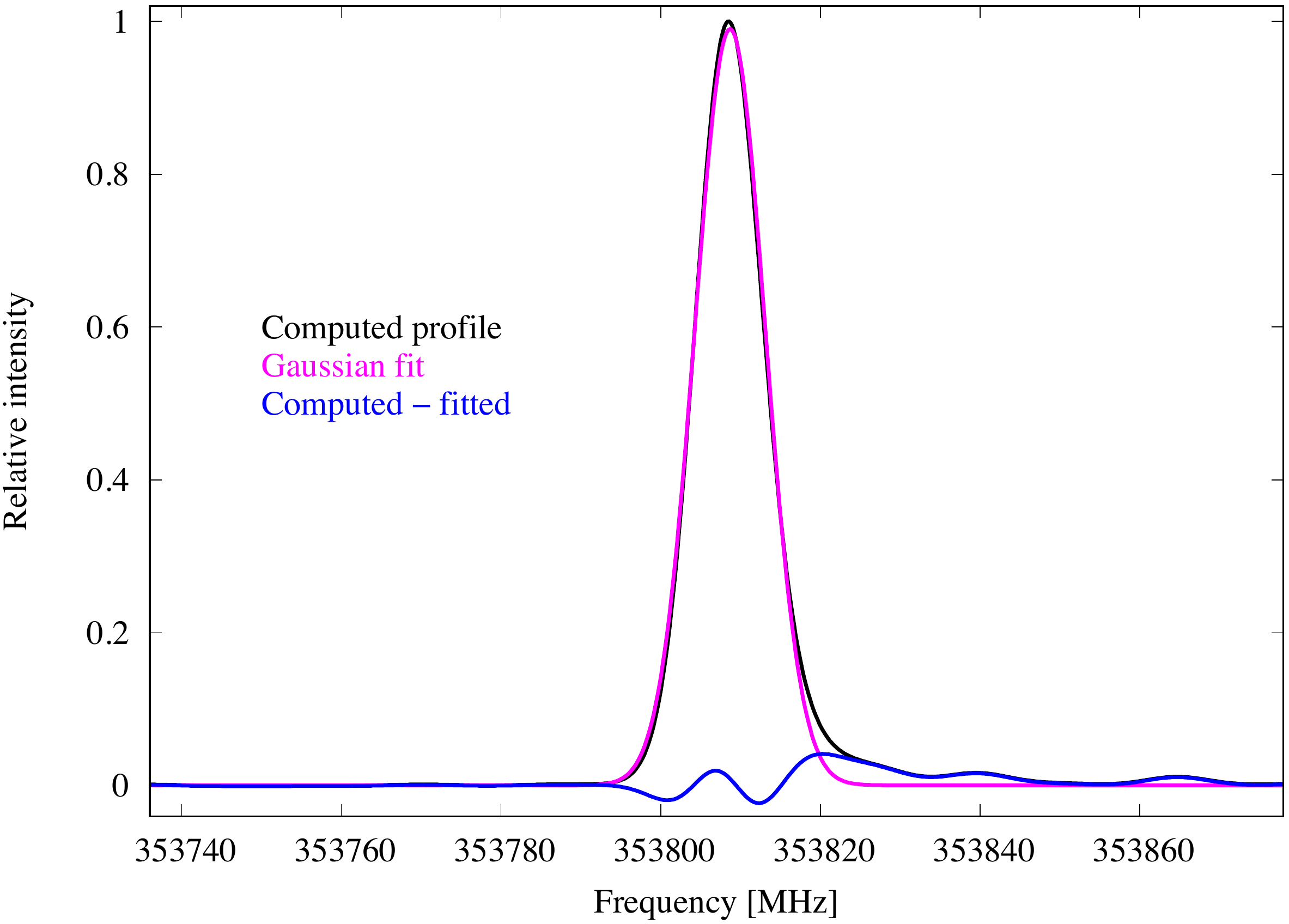}
\caption{Computed model profiles of the 30$\alpha$ (left) and 26$\alpha$ (right) transitions in Fig.~\ref{f:line_blend} are shown in black, together with the best-fitting single Gaussians in cyan. The blue solid line is the difference between them.} 
\label{f:blend_turb_gfit}
\end{figure*}

%
%
\subsubsection{Identification of the line emitters}

Our LTE models further indicate that the observed lines are blends of the corresponding RTLs of Na, Mg (including its three stable isotopes), Al, and Ca, with smaller contributions from Si and Fe if these elements are present in the gas at nearly solar abundances (even though Si has a greater abundance than any of these elements, its contribution to the line emission in strict LTE is smaller, because its ionisation potential is significantly higher); see Table~\ref{t:mod_results}. Mg is the largest contributor, providing 74\,\% of the total intensity. 

We computed theoretical line profiles for the 26$\alpha$ and 30$\alpha$ transitions  for the blend of these eight atomic species (including isotopic variants) in the reference model. The computed model profile represents the weighted sum of the line profiles of the species. Each line component is broadened by a Gaussian of total Doppler width $\Delta \upsilon$. The thermal width is calculated separately for each atom, because it depends upon the atomic mass, while a single value of the micro-turbulent width was assumed for all ($\Delta \upsilon_{\rm turb}$\,=\,7.5\,km\,s$^{-1}$ is adopted so that the fitted FWHMs of the model profiles match the observed width, 8.5\,km\,s$^{-1}$). The computed profiles are displayed in Fig.~\ref{f:line_blend}. The profile shapes of the two transitions are very similar, although the 30$\alpha$ profile is slightly narrower in velocity units than the 26$\alpha$ profile (zero velocity corresponds to the rest frequency of $^{24}$Mg in the simple Rydberg formula). The small shift of the blended profile relative to Mg is quantified in Fig.~\ref{f:blend_turb_gfit} through a least-squares fit of single Gaussian functions to the computed profiles. The fits determine a central rest frequency for the model blends, which are 232\,022.635\,$\pm$\,0.002\,MHz and 353\,808.691\,$\pm$\,0.004\,MHz for the 30$\alpha$ and 26$\alpha$ transitions, respectively. These values agree very well with the rest frequencies of the observed lines estimated assuming a systemic velocity of 41.5\,km\,s$^{-1}$; see Table~\ref{t:obs_rec_res}. The fits also show that the deviations from a Gaussian shape are small, less than 3.0 and 4.2\,\% for the 30$\alpha$ and 26$\alpha$ transitions, respectively.

\subsubsection{Line broadening}
\label{s:line_broadening}

As outlined in Appendix~\ref{a:add_broadening} we can expect line broadening due to fine structure, collisional broadening, radiative damping, and Zeeman splitting in the presence of an external magnetic field, in addition to that of Doppler shifts due to thermal and turbulent motions. 

Unfortunately, there is some uncertainty about the collisional broadening coefficients: adopting the nominal values, the total damping (collisional plus radiative) corresponds to a Lorentzian line shape with FWHMs of $\Gamma$\,=\,2.63 and 3.28\,MHz for 26$\alpha$ and 30$\alpha$, respectively, in the reference model. With this choice, the convolved profile has a Doppler core that can be fit rather well by a Gaussian of $\Delta \upsilon$\,=\,8.5\,km\,s$^{-1}$, as in the observed profile, provided that the turbulent width is $\Delta \upsilon_{\rm turb}$\,=\,5.5\,km\,s$^{-1}$. The deviation from a Gaussian profile is modest, reaching at most 7\,\% of the peak intensity in the blue wing. According to the present theory, the electron-impact collisional broadening has a strong $n$-dependence at higher values of $n$. If this applies also to lower values of $n$, the 30$\alpha$ line is expected to be markedly broader than the 26$\alpha$ line if the density is high enough. This can be used to estimate an upper limit to the density because the observed lines are well fitted with Gaussians of the same width and no indications of Lorentzian wings. Using this (see Appendix~\ref{a:add_broadening}) it appears that we rule out densities higher than $n_{\rm H}$\,$\ga$\,3$\times$10$^{12}$\,cm$^{-3}$, a result independent of the excitation models.

Finally, models with Zeeman splitting have been computed assuming a uniform magnetic field throughout the RTL region. A strong upper limit to the magnetic field of 2\,G can be set because otherwise the 30$\alpha$ line would have a larger line width than the observed one even in the absence of turbulence and damping (and would show a noticeable Zeeman triplet structure). Allowing a combination of turbulence and damping that smooths out the structure of the Zeeman triplet decreases the upper limit on the magnetic field to about 1\,G.

\subsubsection{Dependence on geometry and distance}
\label{s:geom_dist}

An analysis of the sensitivity of the model results to the adopted geometry and distance is presented in Appendix~\ref{a:geom_dist} and summarised in Tables~\ref{t:solutions1} and \ref{t:solutions2}. The primary model solution is not very sensitive to the thickness of the cylinder. As an example of the dependence on geometry, the primary models for $\rho$ between 0.05 and 0.20 are shown in the contour diagram (left panel of Fig.~\ref{f:paraplot}), and they both fall within the hexagon of preferred solutions for $\rho$\,=\,0.1. Even for $\rho$\,=\,2, the temperature and density of the primary solution are not substantially different, $T$\,=\,2650\,K, $n_{\rm H}$\,=\,2$\times$10$^{11}$\,cm$^{-3}$, and are certainly well within the uncertainties of our adopted model. 

The primary model is not very sensitive to the source size either. If we adopt a source size of 17\,mas, that is, the proper size of a uniform brightness disc for which the Gaussian size is 11\,mas,  and $\rho$\,=\,0.1, the resulting temperature and density, 2725\,K and 4$\times$10$^{11}$\,cm$^{-3}$, are only marginally different from those of the reference model, and are well within the uncertainties of our adopted model.  The effect of changing both the source size (17\,mas) and the geometry ($\rho$\,=\,2) is illustrated by the blue rectangle in the left panel of Fig.~\ref{f:paraplot}. We note that a larger source results in lower optical depths for both lines and continuum.

Similarly, we recomputed the models for different values of the distance, $D$\,=\,0.5 and 2\,kpc. The primary solutions are very similar to those for the adopted distance of 1\,kpc because the radius $R$ simply scales to match the observed angular size. Only the gaseous mass depends significantly on the distance. The primary solutions for $\rho$\,=\,0.1 have gas masses of 1.2$\times$10$^{-4}$ and 4.2$\times$10$^{-3}$\,$M_\odot$ for the closer and more distant alternatives, respectively, that is, the gas mass scales roughly as $D^{2.5}$.

%
\subsection{Constraints from optical emission lines}

HD\,101584 has an optical spectrum that is rich in emission and absorption lines. \citet{bakketal96a} made a thorough analysis of their spectra obtained with the ESO Coud{\'e} Auxiliary Telescope, but we note here that they assumed an edge-on geometry as opposed to the face-on geometry advocated by \citet{olofetal19} and \citet{klusetal20}. Here we only select a few emission lines of low optical depth in order to perform a sanity check of the model derived from the radio data.

Numerous emission lines in the visible spectrum can be identified as low-excitation transitions in neutral atoms and a few singly ionised ions. While many of these lines may arise in the chromosphere and inner wind of the star, it is possible that some lines originate in the more extended RTL region that is responsible for the mm-wave line emission. We focused on lines that are most likely to be optically thin, and extracted VLT/UVES data from the ESO archive; see Sect.~\ref{s:opt_obs_desc}. A selection of such lines are listed in Table~\ref{t:opt_lines}. The amplitude, centre, and width of each line was determined from a simultaneous least-squares fit of a low-order polynomial continuum and several Gaussian functions to a small wavelength interval, typically 50\,\AA. The radial velocities, near 41.5\,km\,s$^{-1}$ with respect to the LSR, are in excellent agreement with the systemic velocity of the star and with the derived radial velocities of the X$26\alpha$ and X$30\alpha$ lines at mm wavelengths. However, the observed line widths are significantly larger than those of the mm wave lines, in the range of 15 to 20\,km\,s$^{-1}$. These types of lines were also discussed by \citet{bakketal96a}.

It is interesting to consider whether the RTL-region contributes to any of the observed emission lines in the visible spectrum. Although a full analysis of the visible spectrum is beyond the scope of this paper, a few quantitative comments can be made about. We tried to ensure that the description of the principal continuum opacities applies both in the long-wavelength limit at mm wavelengths and throughout the visible spectrum. At visible wavelengths, both H$^-$ bound-free opacity and Thomson scattering become important compared with the bremsstrahlung that dominates at mm wavelengths. It is re-assuring that the reference model of the RTLs is not in conflict with the intensities of the visible emission lines; see Appendix~\ref{a:opt_lines}.

%
%
%
\subsection{Predictions for radio lines}
\label{s:predictions}

As can be seen in Table~\ref{t:mod_results}, the ionisation degree is low ($\approx$\,10$^{-4}$) in the RTL region and the warm gas consists of predominantly neutral hydrogen, while the molecular fraction is dominated by CO at an abundance of $\approx$\,3\,$\times$\,10$^{-4}$ with respect to hydrogen. The question is whether other atomic or molecular lines can be detected from the RTL region. The results presented in Table~\ref{t:chemistry} show that in the case of molecules, only CO and possibly SiO are viable candidates. Molecular nitrogen, N$_2$, has a high abundance, but its lines are due to weak electric quadrupole transitions.  Predictions are made for the LTE fluxes of some interesting atomic and molecular lines in Table~\ref{t:predictions}.

\begin{table}
 \caption{Predicted radio line flux densities in the reference model\,$^a$}
   \begin{tabular}{lllc}
\hline \hline
Species     & Transition         & Frequency      & $S_{\nu}$  \\
            &                    & [GHz]          & [mJy] \\
\hline  
Mg\,I       & $38\alpha$         & 115.3          & \phantom{0}2.1 \\
Mg\,I       & $54\gamma$         & 115.6          & \phantom{0}0.5 \\
Mg\,I       & $30\alpha$         & 232.0          & \phantom{0}5.9 \\
Mg\,I       & $26\alpha$         & 353.8          & 10\phantom{.0} \\
Mg\,I       & $21\alpha$         & 662.7          & 23\phantom{.0} \\
C\,I        & $^3$P $J=1-0$      & 492.2          & \phantom{0}0.3 \\
CO          & $J=2-1$            & 230.5          & \phantom{0}7.9 \\
CO          & $J=3-2$            & 345.8          & 22\phantom{.0} \\
CO          & $J=4-3$            & 461.0          & 41\phantom{.0} \\
CO          & $v=1, J=2-1$       & 228.4          & \phantom{0}2.5 \\
CO          & $v=1, J=3-2$       & 342.6          & 11\phantom{.0} \\
CO          & $v=1, J=4-3$       & 456.8          & 27\phantom{.0} \\
SiO         & $v=0, J=5-4$       & 217.1          & \phantom{0}5.5 \\ 
SiO         & $v=1, J=5-4$       & 215.6          & \phantom{0}3.5 \\
\hline
\end{tabular}
\label{t:predictions}
\tablefoot{$^{(a)}$ Peak flux densities have been calculated in the same way as for the Rydberg transition lines, with the assumption that the same line width (8.5\,km\,s$^{-1}$) and angular diameter (11\,mas) apply to all.}
\end{table}

In order to compare with our ALMA data, we re-analysed the molecular line data for the CCS component (see Sect.~\ref{s:hd101584}) using only the data obtained with the most extended configuration in Band~6, which therefore has the highest angular resolution, $\approx$\,0\farcs03, which is three times the size of the RTL region.  The S/Ns are high enough to determine the source sizes from the global line profiles, but not high enough to determine their brightness distributions as a function of velocity. We present the results for the CO(\mbox{2--1}) and SiO(\mbox{5--4}) lines because these species have high abundances in the RTL region, and for the H$_2$S(\mbox{$2_{20}-2_{11}$}) and SO$_2$($16_{3,13}-16_{2,14}$) lines because they provide additional kinematic information. The SO$_2$ line comes from a transition with an upper state energy corresponding to 148\,K, that is, a transition at high energy compared to those of the other molecular lines. We also added the upper limit obtained for the CO(\mbox{$\upsilon$\,=\,1, $J$\,=\,3--2}) line at an angular resolution of $\approx$\,0\farcs2. The results are summarised in Table~\ref{t:obs_mol_res}.

\begin{table*}[t]
\caption{Molecular line characteristics of the CCS component.}
\begin{tabular}{l c c c c c c }
\hline \hline
Line                             & $\theta_{\rm b}$\,$^a$           & $\theta_\ell$\,$^b$                         & Aperture        & $S$             & $\Delta \upsilon$\,$^c$    & $\upsilon_{\rm c}$  \\
                                 &  [\arcsec $\times$\arcsec\ (PA)] & [\arcsec $\times$\arcsec\ (PA) ]            & [\arcsec]       & [mJy]           & [km\,s$^{-1}$]             & [km\,s$^{-1}$]       \\
\hline 
CO(2--1)                        & 0.028$\times$0.026 (40$^\circ$)   & 0.180$\times$0.160 (172$^\circ$)              & 0.2\phantom{0}  & 99\phantom{.0}  & 18.7                       & 41.1               \\
                                 &                                  & --                                          & 0.03            & \phantom{0}5.4  & 16.1                       & 43.3 \\
CO($\upsilon$\,=\,1, 3--2)              & 0.22$\times$0.18 (--42$^\circ$)   & --                                          & 0.3\phantom{0}  & <\,5            & --                         & --    \\
SiO(5--4)\,$^d$                  & 0.030$\times$0.028 (36$^\circ$)  & 0.086$\times$0.075 (136$^\circ$)            & 0.1\phantom{0}  &  24\phantom{.0} & 12.3                       & 43.9   \\
                                 &                                  & --                                          & 0.03            & \phantom{0}4.3  & 10.7                       & 44.0 \\
H$_2$S($2_{20}$--$2_{11}$)       & 0.030$\times$0.028 (36$^\circ$)  & 0.088$\times$0.081 (\phantom{0}44$^\circ$)  & 0.1\phantom{0}  & 31\phantom{.0}  & \phantom{0}3.4             & 42.0  \\
                                 &                                  & --                                          & 0.03            & \phantom{0}4.6  & \phantom{0}3.2             & 42.3 \\
SO$_2$($16_{3,13}$--$16_{2,14}$) & 0.030$\times$0.028 (36$^\circ$)  & --                                          & 0.1\phantom{0}  & 32\phantom{.0}  & \phantom{0}1.9             & 41.6   \\
\hline
\end{tabular}
\label{t:obs_mol_res}
\tablefoot{$^{(a)}$ The synthesised beam size and its PA. $^{(b)}$ FWHM of a Gaussian fit to the brightness distribution of the integrated line profile (and deconvolved with the beam). $^{(c)}$ FWHM of a Gaussian fit to the line profile (deconvolved with the spectral resolution of 1.5\,km\,s$^{-1}$).  $^{(d)}$ This line is double-peaked and the results of the Gaussian fit are only indicative.}
\end{table*}

All brightness distributions are larger than the RTL region by about a factor of ten or more (all sizes given here are FWHM of a Gaussian fitted to the brightness distribution and deconvolved with the beam). Except for the SiO line, the lines are reasonably fitted by a Gaussian line profile. For CO, the line width is substantially larger than that of the RTLs, while those of H$_2$S and SO$_2$ are substantially smaller. Only the SiO line has a line width comparable to the RTL line width, but this line is clearly non-Gaussian; it is double-peaked, a result corroborated by the profile of the $^{29}$SiO(\mbox{5--4}) line. Therefore, the source sizes and the line profiles clearly show that the RTL region contributes only marginally, if at all, to the observed molecular lines. Only, the vibrationally excited CO rotational line deviates from this conclusion, but  whether or not LTE is a good approximation for the vibrational excitation of CO is uncertain.

We also estimated the molecular flux densities within an aperture of 0\farcs03 at the centre position. For CO and SiO, these come out close to those estimated for the reference model (Table~\ref{t:predictions}). This suggests that the RTL region may contribute significantly to these observed intensities. In contrast, the H$_2$S and SO$_2$ lines must come from a region well outside the RTL region because the predicted abundances of these molecules are very low in the RTL region (Table~\ref{t:chemistry}) and the lines are very narrow.

With the present data it is not easy to paint a consistent picture of the geometry and kinematics of the different line brightness distributions of the CCS component, including the RTL region. It would seem reasonable that the density and temperature of the gas declines with distance from the centre. This would explain why the more easily excited CO line emission is more extended than those of the other species. Chemistry may also play a role, restricting the three-atom molecules to the higher densities. The nature of the velocity field remains an open question because the observed lines (from mm wave to optical) show a number of different line profiles and line widths.
 
%
%
%
%
\section{Discussion}

\subsection{The origin of the RTL-region}

HD\,101584 is believed to be in the aftermath of a CE process that ended about 800 years ago without a stellar merger \citep{olofetal19}. This resulted in the termination of the RGB evolution of the primary and only its core remains, while the low-mass companion lies at a distance of about 0.5\,au from HD\,101584. The complexity of this process, for which little observational information exists, and the absence of models for stars in this stage make it difficult to speculate about the origin of the warm gas. On the other hand, it does not appear unlikely that about 10$^{-3}$\,$M_\odot$ of warm gas can leave the star at the end of the CE evolution. The gas is most likely stationary ---although small effects of expansion and/or rotation cannot be excluded--- but must be characterised by substantial micro-turbulence (especially if the optical lines discussed above also originate from this region). The exact process responsible for keeping the gas heated and turbulent remains unknown.

%
\subsection{A new tool for studying cooler stars?}

The detection of RTLs from low-ionisation-potential elements towards HD\,101584, combined with the high sensitivity of ALMA, opens up the possibility of also detecting such lines towards other objects of cool to moderate temperature. In particular, red giants on the RGB or AGB may be detectable if they have enough gas in their immediate surroundings with temperatures around 3000\,K and particle densities of the order 10$^{12}$\,cm$^{-3}$. The studies of radio continuum emission from low-ionisation gas in the extended atmospheres of AGB stars indicate that this may very well be the case \citep{reidment07, mattetal18, vlemetal19}. To this category we can also add red supergiants. At temperatures in excess of about 4000\,K we can expect the hydrogen lines to also become detectable. This is, for instance, the case for pre-PNe (without ionising radiation) and objects like yellow supergiants \citep[e.g. \object{Betelgeuse},][]{ogoretal17}  and hypergiants. The H30$\alpha$ line has been reported in ALMA observations of the Fried Egg Nebula \citep{walletal17}, which is associated with the yellow hypergiant star \object{Henize\,3--1379} (\object{IRAS\,17163--3907}).

Close binaries ---like HD\,101584--- of various types are also likely to be of interest in this context. In this category are the symbiotic stars. Several such stars, notably \object{RR~Tel}, show detectable free-free continuum emission at cm wavelengths. An interesting example in the high-mass regime is \object{VV\,Cep}, which is an eclipsing binary star system consisting of a red supergiant that fills its Roche lobe when closest to the companion, a Be star that is most likely on the main sequence \citep{habaetal08}. Here substantial amounts of warm gas can be produced through shocks and the hot radiation field.

The lines due to Rydberg transitions may therefore provide independent information on the warm gas in the vicinity of moderate-temperature stellar objects, and hence provide constraints on the origin and heating of such regions. Rydberg transitions of heavy elements are also susceptible to strong Zeeman splitting, and consequently can be used to estimate stellar magnetic fields, a property that is notoriously difficult to measure for evolved stars \citep[e.g.][]{vlem14}.

%
%
%
%
%
\section{Conclusions}

We identified two lines towards the post-giant star HD\,101584 that can be attributed to the Rydberg transitions 30$\alpha$ and 26$\alpha$ of neutral atoms of elements heavier than carbon. We used a model in strict LTE and a source geometry in the form of a geometrically thin disc to interpret the data and to identify the line carrier(s). A solution that fits all the observational constraints (i.e. the measured line flux densities, upper limits to the continuum flux densities, and the estimated size of the emitting region) includes a gas mass of about 10$^{-3}$\,$M_\odot$, a density of about 10$^{12}$\,cm$^{-3}$, a temperature of about 2800\,K, a size of $\approx$\,10\,au, and a turbulent velocity field in the range 5.5\,--\,7.5\,km\,s$^{-1}$, assuming a distance of 1\,kpc and solar abundances of the elements. The element Mg contributes about 75\,\% of the line fluxes. The contributions from Na, Al, and Ca lie in the range of 5--10\,\% each. The corresponding H and C lines are much weaker. Using the Zeeman effect, we set an upper limit to the average magnetic field in the line-emitting region of 1\,G. 

Further observations are required to determine in more detail the geometry and kinematics of the region where the Rydberg transition lines are formed. The gas appears to be stationary, and the process responsible for keeping the gas heated and turbulent over a region of  about 10\,au remains to be identified; it may be connected to processes active during the CE phase.

Although HD\,101584 is not a standard object, we speculate that the types of lines discussed here should be detectable towards other moderately warm and evolved objects, in particular taking the sensitivity of ALMA into account, and serve as a new tool for their study.

%
%
%
%
\begin{acknowledgements}
We are grateful to the referee (Albert Zijlstra) for providing insightful and constructive comments on the paper. HO and WV acknowledge support from the Swedish Research Council. This paper makes use of the following ALMA data: ADS/JAO.ALMA\#2012.1.00248.S, \#2015.1.00078.S and \#2018.1.00753.S. ALMA is a partnership of ESO (representing its member states), NSF (USA) and NINS (Japan), together with NRC (Canada) and NSC and ASIAA (Taiwan) and KASI (Republic of Korea), in cooperation with the Republic of Chile. The Joint ALMA Observatory is operated by ESO, AUI/NRAO and NAOJ. HO acknowledges support from the Nordic ALMA Regional Centre (ARC) node based at Onsala Space Observatory. The Nordic ARC node is funded through Swedish Research Council grant No 2017-00648. This research has made use of the services of the ESO Science Archive Facility. Based on observations collected at the European Southern Observatory under ESO programme 266.D-5655(A). 
\end{acknowledgements}


\bibpunct{(}{)}{;}{a}{}{,}

%
%
%
%
%

\newpage

\begin{appendix}

%
\section{Elemental abundances}

The abundances are expressed on the standard logarithmic scale [H]\,=\,12.0, so that the relative number densities are $\log(n_{\rm X}/n_{\rm H})$\,=\,$Y({\rm X}) - Y({\rm H}) + \log d({\rm X})$; see Table~\ref{t:abundances}. For Set~1 we adopt solar abundances and all $d({\rm X})$\,$\equiv$\,1.0). The depletion factors for Set~2 are typical of well-mixed, diffuse interstellar gas where refractory elements have already been condensed into dust. Only elements up to Ni with ionisation potentials lower than that of hydrogen are considered.

\begin{table}[h]
 \caption{Adopted elemental abundances}
   \begin{tabular}{lcrl}
   \hline \hline
 X & $I/k_{\rm B}$\,$^a$ & $Y$(X)  & $d$(X)\,$^b$  \\
   &  [K]          &         &   \\
   \hline 
 H   &  157803 & 12.00 &  1.0  \\   
 C   &  130670 &  8.43 &  0.458 \\
 Na  &  \phantom{1}59636 &  6.24  &  0.36 \\
 Mg  &  \phantom{1}88731 &  7.60  &  0.112 \\
 Al  &  \phantom{1}69462 &   6.45  &  0.00179 \\
 Si  &  \phantom{1}94596 &   7.51  &  0.0479   \\
 P  &  121693 &   5.41  &  1.0 \\
 S   &  120223 &  7.12  &  1.0 \\
 Cl  &  150483 &  5.50  &  0.324 \\
 K  &  \phantom{1}50371 &   5.03  &  0.078  \\
 Ca  &  \phantom{1}70940 &  6.34  & 0.0001739 \\
 Ti & \phantom{1}79237 & 4.95 & 0.0022 \\
 Fe  &  \phantom{1}91704 &  7.50  &  0.00375  \\
 Ni  &  \phantom{1}88657 &  6.22  &  0.00176  \\
\hline 
\end{tabular}
\label{t:abundances}
\tablefoot{$^{(a)}$ The ionisation energy ($k_{\rm B}$ is the Boltzmann constant). $^{(b)}$ The depletion factor. Solar photospheric abundances are taken from \citet{aspletal09}. Depleted abundances follow the patterns in diffuse molecular clouds and are taken from \citet{drai11}, except for Na and K \citep{leppetal88}, P \citep{leboetal05, jamealoi18}, and Cl \citep{moometal12}.}
\end{table}

%
%
\section{Chemistry}
\label{a:chemistry}

Table~\ref{t:chemistry} lists the molecules that have been included in the chemistry and ionisation
balance in the LTE models. For each molecule X, the fractional abundance is given with respect to H in the reference model. The adopted dissociation energy $D_0$ is expressed in wavenumber units of cm$^{-1}$ (one electron volt is $\mathrm{eV}$\,=\,8065.5439\,cm$^{-1}$), and the source of the partition function is listed. For most of the included molecules, tables of partition functions were taken from the {\it ExoMol Database} \citep{tennetal16}\footnote{{\tt  http://www.exomol.com/} } and used via interpolation.  In a few cases, notably H$_2$, CO, and N$_2$, the partition functions were calculated directly by summing over states listed in extensive spectroscopic tables compiled for the {\it Leiden Atomic and Molecular Database for Astronomy (LAMDA)} \citep{schoetal05a, vandtaetal20}\footnote{{\tt https://home.strw.leidenuniv.nl/{$\sim$}moldata/}}. 

\begin{table}
 \caption{Molecules and their equilibrium abundances in the reference model}
   \begin{tabular}{lccl}
   \hline \hline
X                 & X/H                  & $D_0$     & Partition function \\
                  &                      & [cm$^{-1}$] & \\
   \hline
H$_2$             & $-\phantom{1}4.34$   & 36118     & state sum \\
H$_2^+$           & $-17.88$             & 21379     & state sum \\
H$_3^+$           & $-19.85$             & 35270     & state sum \\
CO                & $-\phantom{1}3.57$   & 89463     & state sum \\
H$^-$             & $-12.35$             & \phantom{0}6084 & one bound state \\
MgH               & $-12.74$             & 10366     & ExoMol \\
MgO               & $-14.82$             & 21777     & ExoMol \\
SiO               & $-\phantom{1}6.46$   & 66621     & ExoMol \\
SiH               & $-10.29$             & 24358     & ExoMol \\
N$_2$             & $-\phantom{1}4.64$   & 78715     & state sum \\
CN                & $-\phantom{1}9.38$   & 62589     & ExoMol \\
NO                & $-\phantom{1}8.66$   & 52483     & ExoMol \\
SO                & $-11.84$             & 43175     & state sum \\
SO$_2$            & $-19.29$             & 45652     & ExoMol \\
SH                & $-\phantom{1}9.77$   & 29197     & ExoMol \\
TiO               & $-12.46$             & 55488     &  ExoMol \\
TiO$_2$           & $-18.60$             & 50410     & \citet{irwi88} \\
TiO$^+$           & $-11.32$             & 55492     & Black (in prep.)  \\
OH                & $-\phantom{1}7.48$   & 35480     & ExoMol \\
H$_2$O            & $-11.58$             & 41241     & ExoMol \\
HCN               & $-12.33$             & 43247     & ExoMol \\
C$_2$             & $-14.32$             & 50370     & ExoMol \\
CH                & $-11.84$             & 28132     & ExoMol \\
HCO$^+$           & $-17.54$             & 49075     & state sum \\
TiH               & $-15.80$             & 16776     & ExoMol \\
H$_2$S            & $-15.84$             & 30938     & ExoMol \\
FeH               & $-12.66$             & 12824     & ExoMol \\
FeO               & $-13.15$             & 33715     & ExoMol \\
CS                & $-11.12$             & 59322     & ExoMol \\
CO$_2$            & $-11.57$             & 44399     & ExoMol \\
\hline
\end{tabular}
\label{t:chemistry}
\end{table}

Although 30 diatomic and triatomic molecules have been included in the equilibrium computation of ionisation, only a few of them have significant abundances. Notice that the primary solutions reflect conditions in which a small fraction of the hydrogen is in molecular form, while a significant fraction of the total carbon and oxygen is consumed in CO. The relatively high abundances of CO, N$_2$, and SiO reflect their high dissociation energies. At $T$\,$\approx$\,2800\,K and $n_{\rm H}$\,$\approx$\,10$^{12}$\,cm$^{-3}$ in a gas of solar abundances, the presence of these most abundant molecules does have a small effect on the overall ionisation balance and thus the intensity of the free-free continuum emission. In addition, these models can be used to predict the intensity of molecular line emission that should accompany the atomic lines under the adopted assumptions about strict LTE and solar abundances.

%
%
%
\section{Additional line broadening} 
\label{a:add_broadening}

Because Rydberg atoms interact strongly with electrons, other atoms, radiation, and magnetic fields, it is worthwhile exploring whether or not the line profiles contain additional information about the conditions in the emitting regions. This can be done through computation of theoretical line profiles based on the conditions in the reference model described above. In order to do this properly, the fine structure of the Rydberg transitions will have to be added.

\subsection{Fine structure}

Rydberg states of principal quantum number $n$ possess slightly non-degenerate fine structure in their orbital substates (quantum number $\ell$\,=\,0 to $n-1$). Hence, a Rydberg transition $n'$\,$\to$\,$n''$ is composed of many fine-structure components $(n',\ell')$\,$\to$\,$(n'',\ell'')$ that obey the selection rules $\ell'$\,=\,$\ell''$\,$\pm$\,1. Although the fine-structure splitting is negligible at $n$\,=\,25 to 30 in hydrogen \citep{towletal96}, atoms like Mg, Al, and Ca have parental ionic cores with rather high values of the static dipole polarisability, which gives rise to larger fine-structure splittings than in hydrogen. Thus, the $26\alpha$ transition of each atomic emitter has 51 fine-structure components and the model profile for this transition is a blend of 408 lines for the atomic species considered in the reference model; see Table~\ref{t:blend}. The same figures for the $30\alpha$ transition are 59 and 472. In the construction of a theoretical profile, each of the fine-structure components is weighted by the relative abundance of the atomic species and by an intensity factor, which is proportional to the upper-state statistical weight multiplied by the spontaneous transition probability.

We computed the energy of each state of each atom, labelled $(n,\ell)$, from the solution of the Dirac equation for a hydrogenic atom with $n$\,$\gg$\,1 as presented in equations 2 through 4 of \citet{towletal96}. An additional term, $-\alpha_{\rm d} P(n,\ell)$, is included to account for the polarisation of the ionic core, where $P(n,\ell)$ is a function of $n$ and $\ell$, and $\alpha_{\rm d}$ is the electric dipole polarisability \citep{channoye83, chan87}. In H itself, with a bare proton core, the polarisability vanishes and this term is zero. In Mg, on the other hand, the Mg$^+$ core has $\alpha_{\rm d}$\,=\,33.05$\pm$0.03\,$a_0^3$ (atomic units, where $a_0$ is the Bohr radius \citep[see][]{mitretal10}). We computed an $\ell$-resolved line list for each transition $n'$\,$\to$\,$n''$. The number of fine-structure components (distinct lines) in an $\alpha$ transition is $2n''$--1, where double-prime indicates the lower state. The spontaneous transition probabilities (Einstein $A$-coefficients) for these hydrogenic transitions were computed through use of the computer program of \citet{storhumm91}. We assume that pure L-S coupling applies. For Rydberg states in the limit of high-$n$, all atoms are assumed to behave like degenerate hydrogenic spin-doublets of total angular momentum $j$\,=\,$\ell$$\pm$1/2. The degeneracy is lifted in the presence of external fields, as we explain below in the section on Zeeman splitting. In the absence of external fields, the blending of lines of different atoms and the overlap of their fine-structure components broaden the line profile and make it slightly asymmetrical. The fine-structure splittings in the 26$\alpha$ and 30$\alpha$ transitions correspond to only a few tenths of a km\,s$^{-1}$, and hence will not contribute substantially to the line widths measured here.

\subsection{Collisional broadening}

Atoms in Rydberg states interact with electrons and neutral H atoms through elastic and inelastic collisions that perturb their energies and thus broaden their line profiles. This collisional broadening depends on the densities of the collision partners and on the kinetic temperature. In ionised nebulae, recombination lines suffer collisional broadening mainly with electrons and ions. In contrast, our reference model is only weakly ionised, with $n_{\rm e}/n_{\rm H}$\,=\,3.4$\times$10$^{-5}$; as a result, neutral hydrogen and electrons are the most important collision partners. 

The electron-impact broadening of nebular recombination lines has been discussed in the literature; unfortunately, most of that work is not applicable here. \citet{wats06} presented a formula for the collisional broadening based on the classic work of \citet{grie67}. As pointed out by \citet{alexgula16}, that formula contains errors. Beyond that, Griem's approximation was originally intended for cm- and dm-wave transitions with $n$\,>\,100; the approximation breaks down when $n$\,$<$\,70 and is thus not appropriate for the transitions of interest here. \citet{brocleem71} computed the electron--hydrogen cross-sections and rates numerically, and evaluated the line-broadening without the restrictive assumptions of semi-classical theory. Their results could be approximated by a simple fitting formula for
the collisional line width of $\alpha$ transitions,
\begin{equation}
\frac{\Delta \nu_{\rm e}}{\nu_0} = 1.43\times10^{-5} \left( \frac{n}{100} \right)^{7.4}\, \left( \frac{10^4\,{\rm K}}{T} \right)^{0.1} \left( \frac{n_{\rm e}}{10^4\,{\rm cm}^{-3}} \right) 
,\end{equation}
where $T$ and $n_{\rm e}$ are the temperature and number density of electrons, respectively
\citep{brocseat72}. Because the original computations did not extend to $n$\,$<$\,109 and because the dependence on $n$ is so steep, it is not certain that this formula is accurate at $n$\,=\,26 and 30. If we simply apply the formula for $T$\,=\,2805\,K and $n_{\rm e}$\,=\,2.8$\times$10$^7$\,cm$^{-3}$, as in the 
reference model, then the electron-impact widths (FWHM) are $\Delta \upsilon_{\rm e}$\,=\,0.63 and 1.83\,km\,s$^{-1}$ for $n$\,=\,26 and 30, respectively. We note that the collisional broadening produces a Lorentzian line shape rather than a Gaussian. Moreover, if electron-impact broadening were to dominate, the width of the $30\alpha$ line would be almost three times larger than that of $26\alpha$ in velocity units according to the formula of \citet{brocseat72}. 

Collisional broadening by neutral atoms has also been considered, particularly in connection with laboratory experiments on Rydberg atoms in gases of noble or alkali atoms \citep{omon77, kaul84}. Neutral--atom collisions are also important in line-broadening in stellar atmospheres. Specifically, \citet{hoanvanr95} applied the impulse approximation to compute the broadening of mid-infrared lines of Mg\,I by atomic hydrogen in the photosphere of the Sun, where these lines are observed in emission
at wavelengths near 12\,$\mu$m. These are also Rydberg transitions, but at the relatively low values of the principal quantum number involved,  $n$\,=\,5, 6, and 7, the orbital and spin-orbit fine structures are well resolved.  If we adapt the theoretical widths calculated by \citet{omon77} to the case of H on Mg, we estimate $\Delta \nu_{\rm H}$\,$\approx$\,1.8\,MHz at both $n$\,=\,26 and 30, corresponding to velocity widths $\Delta \upsilon_{\rm H}$\,=\,1.5 and 2.2\,km\,s$^{-1}$, respectively, in the reference model. Even though the neutral--neutral interactions are much weaker than electron--neutral interactions involving Rydberg states, the estimated contributions to the line broadening are comparable in our case because of the low electron fraction. 

In both the cases of electron and neutral collisions, the collisional broadening should be re-computed numerically with modern methods, because both elastic and inelastic collisions are important and earlier approximations at $n$\,$>$\,100 are known to break down when $n$\,$\sim$\,20 to 30, but this is beyond the scope of this paper. The above estimates suggest that the collisional broadening might become noticeable if the line profile can be observed with sufficiently high S/N to reveal Lorentzian wings. However, the existing data show line profiles that are fit very well by single-Gaussian functions, meaning that, even with the uncertainties in the theoretical line widths, the current data on HD\,101584 suggest that the neutral density in the emitting region cannot be as large as $10^{13}$\,cm$^{-3}$, otherwise the Lorentzian wings would be apparent. This kind of limit on density is not model dependent; moreover, it could be refined with improved theory.

\subsection{Radiation damping}

Atoms in Rydberg states also interact with electromagnetic radiation. The natural line broadening (radiation damping) via spontaneous transitions can be further enhanced through stimulated emission and absorption in the ambient radiation field. Considering a transition from upper state u to lower state l, the energies of these states are diffuse because both states have finite lifetimes. The Lorentzian FWHM of the radiation damping is $\Delta\nu_{\rm rad}$\,=\,$\gamma/(2\pi)$ where 
\begin{eqnarray}
\gamma_{\rm ul} & = & \sum_{{\rm i}<{\rm u}} \left( A_{\rm ui} + {\bar J}_{\rm ui} B_{{\rm u} {\rm i}} \right) + \sum_{{\rm k}>{\rm u}}  {\bar J}_{\rm ku} B_{\rm ku}  + 
\sum_{{\rm i}'<{\rm l}} \left( A_{\rm li'} + {\bar J}_{\rm li'} B_{\rm li'} \right) + \nonumber \\
    & & + \sum_{{\rm k}'>{\rm l}}  {\bar J}_{\rm k'l} B_{\rm k'l} \,\, ,
\end{eqnarray}
and $A_{\rm ik}$ is the spontaneous transition probability for transition i\,$\to$\,k, and ${\bar J}_{\rm ik}$ the angle-averaged intensity of radiation at the frequency $\nu_{\rm ik}$ of the transition.  The Einstein coefficients for stimulated emission, $B_{\rm ik}$\,=\,$A_{\rm ik}$/(2h$\nu^3$/$c^2)$, and absorption, $B_{\rm ki}$\,=\,$g_{\rm i} B_{\rm ik}$/$g_{\rm k}$ are related to $A_{\rm ik}$ and to the statistical weights of the upper and lower states $g_{\rm i}$ and $g_{\rm k}$. When the radiation can be represented by a blackbody diluted by a geometrical factor $w$\,$\leq$\,1, then
\begin{equation}
{\bar J}_{\rm ik} = w B_{\nu_{\rm ik}}(T_{\rm rad})  = \frac{w 2 h{\nu_{\rm ik}^3}/c^2}{\exp\left(h\nu_{\rm ik}/k_{\rm B} T_{\rm rad}\right) - 1} \,\, ,
\end{equation}
in terms of the Planck function $B_{\nu}$ at temperature $T_{\rm rad}$. Thus, the damping coefficient can be re-written
\begin{eqnarray}
\gamma_{\rm ul} & = & \sum_{{\rm i}<{\rm u}} A_{\rm ui}\left( 1 + b(\nu_{\rm ui}) \right) +
 \sum_{{\rm k}>{\rm u}}  A_{\rm ku} b(\nu_{\rm ku}) \frac{g_{\rm k}}{g_{\rm u}} \nonumber \\
   & & + \sum_{{\rm i}'<{\rm l}} A_{\rm li'} \left( 1 + b(\nu_{\rm li'}) \right) +  \sum_{{\rm k}'>{\rm l}}  A_{\rm k'l} b(\nu_{\rm k'l}) \frac{g_{\rm k}'}{g_{\rm l}}
,\end{eqnarray}
where $b(\nu)$\,=\,$w$/$\left(\exp(h\nu/k_{\rm B} T_{\rm rad}) - 1\right)$. Our model of the line-emitting region assumes strict LTE at the kinetic temperature $T$. This means that $w$\,=\,1 and $T_{\rm rad}$\,=\,$T$ should be a good approximation at frequencies less than that of the $26\alpha$ transition, owing to the inverse frequency dependence of the predominant opacity source. However, some of the stimulated transitions occur at much higher frequencies where the emitting region may be transparent to radiation from the much hotter visible photosphere of HD\,101584, for which $w$\,$<$\,1 and $T_{\rm rad}$\,$\sim$\,8500\,K. 

The computation of  $\Delta\nu_{\rm rad}$ is straightforward for hydrogenic transition frequencies and transition probabilities and is done here by summing all rates for states up to $n$\,=\,1000. The radiation damping of the 26$\alpha$ and 30$\alpha$ lines at $T_{\rm rad}$\,=\,2805\,K and $w$\,=\,1 gives $\Delta\nu_{\rm rad}$ corresponding to 0.072 and 0.083\,km\,s$^{-1}$\,km\,s$^{-1}$, respectively. Even in the case of the hot photosphere $T_{\rm rad}$\,=\,8500\,K shining through the emission region at $w$\,=\,1/2, the damping widths would be increased by only about 50\,\%. In either case, the radiation damping is negligible in comparison with other broadening mechanisms. In general, this broadening effect on Rydberg transitions might be most important at very low radio frequencies ($n$\,$>$\,200) in gas exposed to intense non-thermal continuum radiation, as in the narrow-line regions of quasars \citep[cf.][]{wadietal83}.

Both the collisional line broadening and the radiation damping produce Lorentzian profiles. Two Lorentz functions combine to form a new Lorentz function whose width is simply the sum of the first two Lorentz widths. The total profile is obtained by a convolution of the final Lorentz function with the profile for the blend.

\subsection{Magnetic fields and Zeeman broadening}

Magnetic fields can cause a line broadening in radio and mm-wave RTLs through the Zeeman effect. The blending of lines of different emitting atoms, each with the orbital fine structure described above, is further complicated in the presence of external fields. A one-electron, hydrogenic state with spin $s$\,=\,1/2, labelled by the quantum numbers $n$, $\ell$, and $j$ in L-S coupling, has a pair of degenerate levels for each $j$\,=\,$\ell$\,$\pm$\,1/2. The degeneracy, that is, the number of unresolved sublevels, is 2$j$+1. A magnetic field of strength $B$ lifts this degeneracy, and yields splittings in the weak-field limit of the Zeeman effect given by 
\begin{equation}
E(n,\ell,j,m_j) = E(n,l) + g \mu_{\rm B} B m_j 
,\end{equation}
where $m_j$\,=\,-$j$, -$j$+1, $\dots$, +$j$-1, +$j$. In L-S coupling the Land{\'e} $g$-factor is 
\begin{equation}
g = 1 + \frac{j(j+1) - \ell(\ell + 1) + s(s+1)}{2j(j+1)}
,\end{equation}
and $g$\,$\approx$\,1 for the levels of high $j$ and $\ell$ that contribute most to the line profile. We note that $\mu_{\rm B} B/h$\,=\,1.3996\,$B$\,MHz ($B$ is in the unit of gauss), so that a field strength of the order of $B$\,$\approx$\,1\,G can produce a splitting of the order of 1\,MHz, which may be significant in the lines of interest here.

\subsection{The blended line profile}

We have computed theoretical line profiles for the 26$\alpha$ and 30$\alpha$ transitions taking into account all of the effects of blending and broadening. Eight atomic species are included, as listed in Table~\ref{t:blend}. The computed model profile represents the weighted sum of the scaled optical depths of the various $(n,\ell)$ line components. Each line component is broadened by a Gaussian of total Doppler width $\Delta \upsilon$. The thermal width is calculated separately for each atom, because it depends upon the atomic mass, while a single value of the micro-turbulent width was assumed for all.

\begin{table}
 \caption{Blending and fine structure}
   \begin{tabular}{lcllll}
   \hline \hline
   Atom & Intensity\,$^a$ & \multicolumn{4}{c}{Line components} \\  
        & weighting & \multicolumn{2}{c}{26$\alpha$} & \multicolumn{2}{c}{30$\alpha$} \\
                   &             & $n,\ell$ & $n,\ell,j,m_j$ & $n,\ell$ & $n,\ell,j,m_j$ \\
\hline
 $^{24}$Mg & 0.5825 & \phantom{0}51 & \phantom{0}3900 & \phantom{0}59 & \phantom{0}5220 \\
 $^{25}$Mg & 0.0737 & \phantom{0}51 & \phantom{0}3900 & \phantom{0}59 & \phantom{0}5220 \\
 $^{26}$Mg & 0.0812 & \phantom{0}51 & \phantom{0}3900 & \phantom{0}59 & \phantom{0}5220 \\
             Na & 0.0662 & \phantom{0}51 & \phantom{0}3900 & \phantom{0}59 & \phantom{0}5220 \\
             Al  & 0.1060 & \phantom{0}51 & \phantom{0}3900 & \phantom{0}59 & \phantom{0}5220 \\
             Ca & 0.0832 & \phantom{0}51 & \phantom{0}3900 & \phantom{0}59 & \phantom{0}5220 \\
             Si & 0.0458 & \phantom{0}51 & \phantom{0}3900 & \phantom{0}59 & \phantom{0}5220 \\
             Fe & 0.0189 & \phantom{0}51 & \phantom{0}3900 & \phantom{0}59 & \phantom{0}5220 \\
 Total     & & 408 & 31200 & 472 & 41760 \\         
\hline
\end{tabular}
\label{t:blend}
\tablefoot{(a) The weighting factors reflect the equilibrium abundances and level populations of the neutral atoms in the reference model. Magnesium isotopes are assumed to be present in the same relative amounts as in standard terrestrial material, (0.79:0.10:0.11) for isotopic masses (24:25:26). Columns 3-6 list the numbers of fine-structure components, without and with an external magnetic field. These are summed in the last line of the table.}
\end{table}

Further, we computed a series of model profiles with several values of the damping width, $\Gamma$, the FWHM of the Lorentzian function that describes the combined effect of collisional broadening and radiation damping. The total profile is a convolution of the blend with a Lorentzian function of width $\Gamma$. The collisional broadening term scales directly with density. As discussed above, there is some uncertainty about the collisional broadening coefficients: if we adopt the nominal values, then the total damping is $\Gamma$\,=\,2.63 and 3.28\,MHz for 26$\alpha$ and 30$\alpha$, respectively, for the reference model. With this choice, the convolved profile has a Doppler core that can be fit rather well by a Gaussian of $\Delta \upsilon$\,=\,8.5\,km\,s$^{-1}$, like the observed profile, provided that the turbulent width is $\Delta \upsilon_{\rm turb}$\,=\,5.5\,km\,s$^{-1}$ (in contrast to the best model with no damping wings that requires $\Delta \upsilon_{\rm turb}$\,=\,7.5\,km\,s$^{-1}$). The deviation from a Gaussian profile is modest, and reaches about 7\,\% of the peak intensity in the blue wing. It therefore appears that the collisional broadening effect will not be conspicuous in the analysis of  only one line. 

\begin{figure}
\centering
\includegraphics[width=8cm]{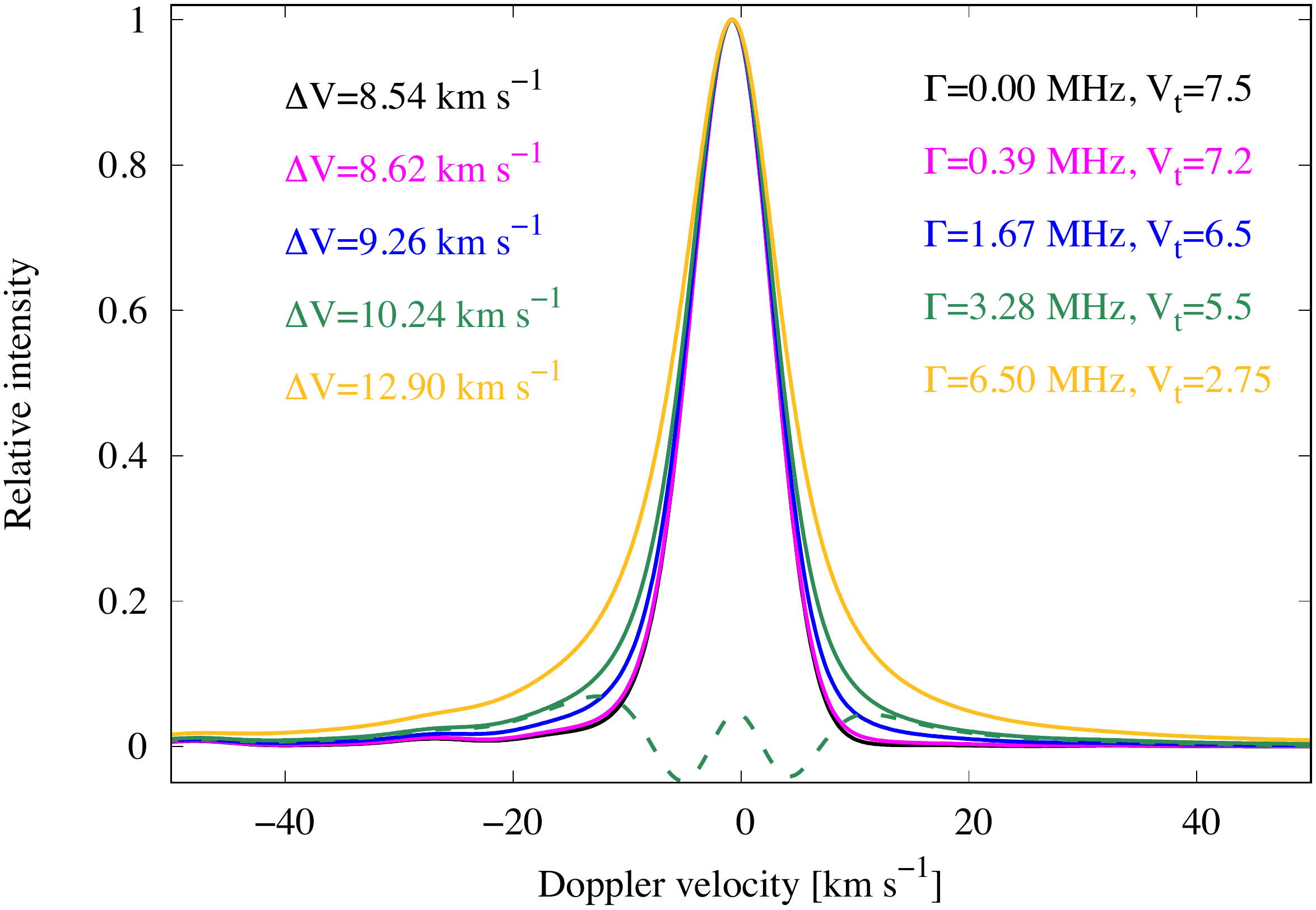}
\caption{Series of model profiles of the 30$\alpha$ line for different values of the Lorentzian width $\Gamma$ and turbulent widths that give $\Delta \upsilon$\,=\,8.5\,km\,s$^{-1}$ in the 26$\alpha$ line. The fitted Gaussian widths are indicated: they exceed the observed width when $\Gamma$\,$>$\,1.7\,MHz. The deviation from a single Gaussian is shown for the nominal case as a dashed green curve.}
\label{f:30_blended_coll} 
\end{figure}

The situation becomes more interesting if the electron-impact collisional broadening does indeed have the strong $n$-dependence predicted by the existing theory at higher values of $n$. In that case, the best-fitting Gaussian width of the 30$\alpha$ transition is expected to be somewhat different from that of the 26$\alpha$ transition. Figure~\ref{f:30_blended_coll} shows the corresponding 30$\alpha$ profiles for the conditions that yield effective line widths of 8.5\,km\,s$^{-1}$ in the 26$\alpha$ line. Here, the nominal collisional broadening of the 30$\alpha$ line gives a profile with a best-fitting Gaussian width that is clearly larger, $\Delta \upsilon$\,=\,10.2\,km\,s$^{-1}$. The collisional broadening width would need to be a factor of two times smaller in order for the effective widths of the two transitions to be approximately the same, if both lines are truly formed at the same density and temperature and turbulent width. This factor of two is within the uncertainty of the extrapolated theory of collisional broadening. On the other hand, collisional broadening two times larger than the nominal case would yield noticeable Lorentzian wings and a 50\,\% difference in fitted FWHM between the two transitions. These model profiles were computed with reference to an LTE excitation model with density $n_{\rm H}$\,$\approx$\,8$\times$10$^{11}$\,cm$^{-3}$. Thus it appears that the observed line profiles can already rule out densities higher than $n_{\rm H}$\,$\ga$\,3$\times$10$^{12}$\,cm$^{-3}$, independent of excitation models.

\begin{figure}
\centering
\includegraphics[width=8cm]{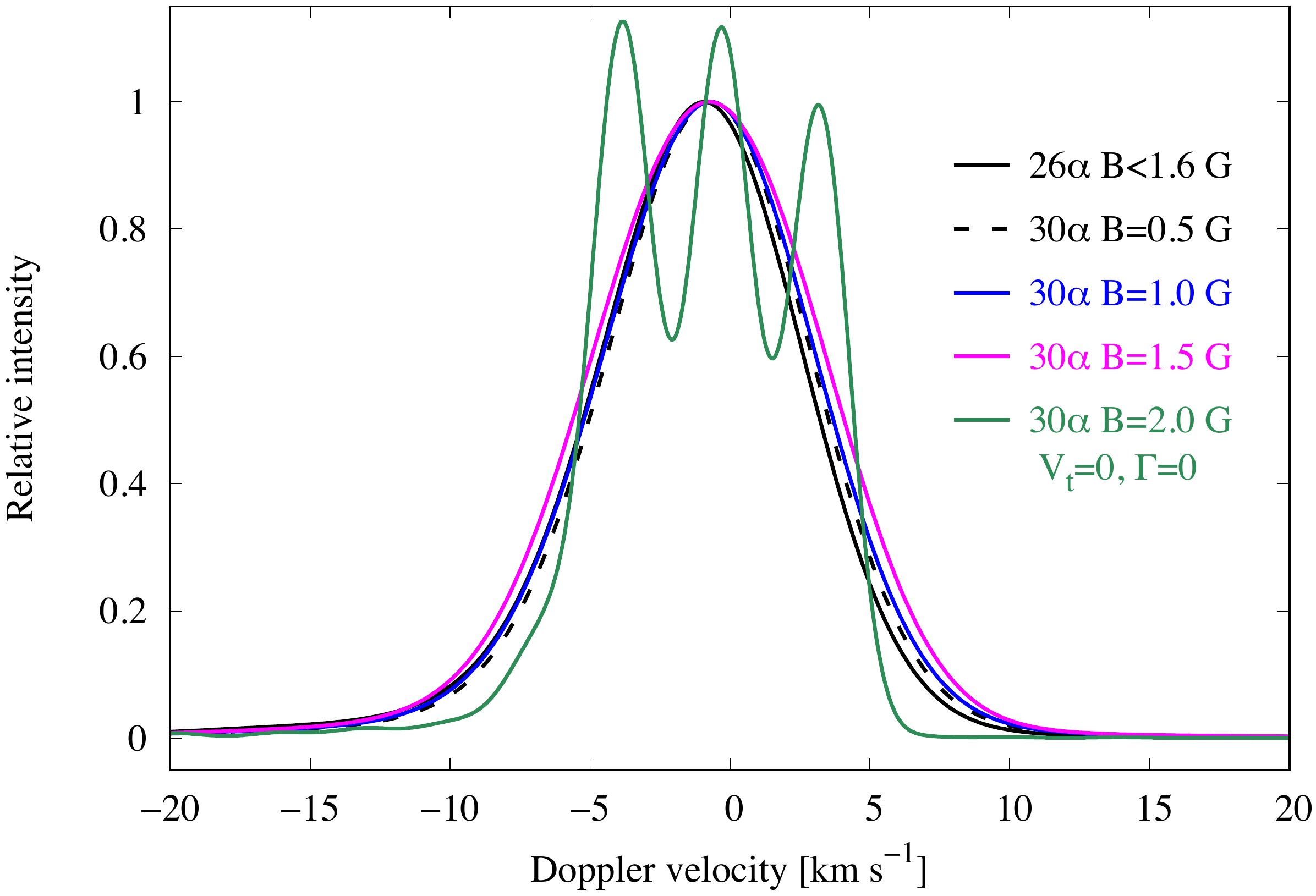}
\caption{Series of model profiles including Zeeman splitting. Examples with small damping ($\Gamma$\,=\,0.34\,MHz in 26$\alpha$) are shown in black. The structure of the 26$\alpha$ profile is smoothed by turbulence and damping to approximate a Gaussian shape with $B$ in the range 0.5 to 1.6\,G. The corresponding 30$\alpha$ profile is measurably broader, $>$\,9.0\,km\,s$^{-1}$ when $B$\,$>$1\,G. A strong upper limit of $B$\,$<$\,2\,G is implied by the fact that the 30$\alpha$ width is too 
large even when turbulence and damping are turned off, which then leaves the 
Zeeman triplet resolved (green curve).}
\label{f:26_30_blended_zeeman} 
\end{figure}

Finally, we have computed models with fully expressed Zeeman splitting for several combinations of  magnetic field strength $B$ (assumed uniform through the source), turbulent width $\Delta \upsilon_{\rm turb}$, and damping width $\Gamma$. A few examples of Zeeman-broadened profiles are shown in Fig.~\ref{f:26_30_blended_zeeman}. The combination of turbulence and damping tends to smooth out the structure of the Zeeman triplet when $B$\,$<$\,2\,G, so that the 26$\alpha$ profiles can all be made to approach Gaussian shapes with $\Delta \upsilon$\,=\,8.5\,km\,s$^{-1}$ when the damping is small and $B$\,$\leq$\,1.5\,G). However, the corresponding 30$\alpha$ profiles have Gaussian-fitted widths $\Delta \upsilon$\,$\geq$\,9.0\,km\,s$^{-1}$ when $B$\,$\geq$\,1.0\,G. As the observed line widths of the two transitions are approximately equal, this suggests an upper limit to the field strength of 1\,G when the damping is small. The observed width of the 30$\alpha$ transition sets a strong upper limit of $B$\,$<$\,2\,G, because the fitted width of the computed profile is 9.2\,km\,s$^{-1}$ even when $\Delta \upsilon_{\rm turb}$\,=\,0 and $\Gamma$\,=\,0 at $B$\,=\,2\,G. Again, the strong limit on magnetic field is independent of any excitation model, and is insensitive to uncertainties in the collisional broadening theory, but it depends on the field geometry. 

%
%
\section{Dependences on geometry and distance}
\label{a:geom_dist}

Additional models were computed by us in order to explore the sensitivity of the solutions to the geometrical parameter $\rho$ and the distance $D$. We considered values of $\rho$ from 0.05 to 2.0 and distance ranges from 0.5 to 2.0\,kpc. All models were constrained to fit the observed results at 232\,GHz:  $\theta$\,=\,11.0\,mas, $S_{\nu}({\rm X}30\alpha)$\,=\,7.3\,mJy, and $S_{\rm c}(232)$\,=\,7.5\,mJy. Details of additional models are summarised in Tables~\ref{t:solutions1} and \ref{t:solutions2}.

The models all behave similarly and the differences in best-fitting temperature (within 100\,K) and total density (within a factor of 3) are relatively small. The primary solutions all have continuum optical depths of the order of unity at 232\,GHz, and the X30$\alpha$ line optical depths for each contributing element are small enough that simply combining them in the optically thin limit is a good approximation.

\begin{table}
 \caption{Models reflecting a larger range of parameter space\,$^a$.}
   \begin{tabular}{lll}
   \hline \hline
   Property & & Primary solution \\
   \hline
   $D$ & [kpc] & 1.0 \\
   $\rho$ & & 0.1  \\
   $T$ & [K] &  2805  \\
   $n(\rm H)$ & [cm$^{-3}$] & $8.2\times 10^{11}$ \\
   $R$ & [au] & 5.5  \\
   $\tau_{\ell}$(232\,GHz) & & 1.2 \\
   $\tau_{\rm c}$(232\,GHz) & & 1.3 \\
   $n(e)$ & [cm$^{-3}$] & $2.8\times 10^7$  \\
   $M_{\rm gas}$ & [M$_{\odot}$] & $7.2\times 10^{-4}$  \\
   $n({\rm H}_2)$ & [cm$^{-3}$] & $3.7\times 10^7$ \\
   $n({\rm CO})$ & [cm$^{-3}$] &  $2.2\times 10^8$  \\
   $n({\rm OH})$  & [cm$^{-3}$] &  $2.7\times 10^4$  \\
   $n({\rm SiO})$  & [cm$^{-3}$] &  $2.8\times 10^5$ \\
   $n({\rm N}_2)$ & [cm$^{-3}$] & $1.9\times 10^7$  \\
   \hline
   $D$ & [kpc] & 1.0  \\
   $\rho$ & & 0.05 \\
   $T$ & [K] &  2845  \\
   $n(\rm H)$ & [cm$^{-3}$] & $1.1\times 10^{12}$ \\
   $R$ & [au] & 5.5 \\
   $\tau_{\ell}$(232\,GHz) & & 1.2 \\
   $\tau_{\rm c}$(232\,GHz) & & 1.3 \\
   $n(e)$ & [cm$^{-3}$] & $3.9\times 10^7$  \\
   $M_{\rm gas}$ & [M$_{\odot}$] & $4.6\times 10^{-4}$  \\
   $n({\rm H}_2)$ & [cm$^{-3}$] & $5.2\times 10^7$  \\
   $n({\rm CO})$ & [cm$^{-3}$] & $3.0\times 10^8$  \\
   $n({\rm OH})$ & [cm$^{-3}$] & $4.1\times 10^4$  \\
   $n({\rm SiO})$ & [cm$^{-3}$] & $3.4\times 10^5$  \\
   $n({\rm N}_2)$ & [cm$^{-3}$] & $2.5\times 10^7$  \\
   \hline
   $D$ & [kpc] & 1.0  \\
   $\rho$ & & 0.2   \\
   $T$ & [K] &  2770  \\
   $n(\rm H)$ & [cm$^{-3}$] & $6.0\times 10^{11}$  \\
   $R$ & [au] & 5.5  \\
   $\tau_{\ell}$(232\,GHz) & & 1.3 \\
   $\tau_{\rm c}$(232\,GHz) & & 1.3  \\
   $n(e)$ & [cm$^{-3}$] & $2.0\times 10^7$  \\
   $M_{\rm gas}$ & [M$_{\odot}$] & $1.2\times 10^{-3}$  \\
   $n({\rm H}_2)$ & [cm$^{-3}$] & $2.5\times 10^7$  \\
   $n({\rm CO})$ & [cm$^{-3}$] & $1.6\times 10^8$  \\
   $n({\rm OH})$ & [cm$^{-3}$] & $1.8\times 10^4$  \\
   $n({\rm SiO})$ & [cm$^{-3}$] & $2.3\times 10^5$  \\
   $n({\rm N}_2)$ & [cm$^{-3}$] & $1.4\times 10^7$  \\
   \hline
   $D$ & [kpc] & 1.0  \\
   $\rho$ & & 1.0   \\
   $T$ & [K] &  2690  \\
   $n(\rm H)$ & [cm$^{-3}$] & $2.9\times 10^{11}$  \\
   $R$ & [au] & 5.5  \\
   $\tau_{\ell}$(232\,GHz) & & 1.4 \\
   $\tau_{\rm c}$(232\,GHz) & & 1.4  \\
   $n(e)$ & [cm$^{-3}$] & $8.6\times 10^6$  \\
   $M_{\rm gas}$ & [M$_{\odot}$] & $5.7\times 10^{-4}$  \\
   $n({\rm H}_2)$ & [cm$^{-3}$] & $1.0\times 10^7$  \\
   $n({\rm CO})$ & [cm$^{-3}$] & $7.8\times 10^7$  \\
   $n({\rm OH})$ & [cm$^{-3}$] & $7.4\times 10^3$  \\
   $n({\rm SiO})$ & [cm$^{-3}$] & $1.5\times 10^5$  \\
   $n({\rm N}_2)$ & [cm$^{-3}$] & $7.4\times 10^6$  \\
   \hline
\end{tabular}
\label{t:solutions1}
\tablefoot{(a) All of the above models satisfy the observed constraints at 232\,GHz:  $\theta$\,=\,11\,mas, $S_{\nu}(30\alpha)$\,=\,7.3\,mJy, and and $S_{\rm c}(232)$\,=\,7.5\,mJy.} 
\end{table}

\begin{table}
 \caption{Models reflecting a larger range of parameter space, continued.}
   \begin{tabular}{lll}
\hline \hline   
   Property & & Primary solution  \\
\hline
   $D$ & [kpc] & 1.0  \\
   $\rho$ & & 2.0   \\
   $T$ & [K] &  2654  \\
   $n(\rm H)$ & [cm$^{-3}$] & $2.1\times 10^{11}$  \\
   $R$ & [au] & 5.5  \\
   $\tau_{\ell}$(232\,GHz) & & 1.4 \\
   $\tau_{\rm c}$(232\,GHz) & & 1.4  \\
   $n(e)$ & [cm$^{-3}$] & $6.0\times 10^6$  \\
   $M_{\rm gas}$ & [M$_{\odot}$] & $4.2\times 10^{-4}$  \\
   $n({\rm H}_2)$ & [cm$^{-3}$] & $7.3\times 10^6$  \\
   $n({\rm CO})$ & [cm$^{-3}$] & $5.7\times 10^7$  \\
   $n({\rm OH})$ & [cm$^{-3}$] & $5.1\times 10^3$  \\
   $n({\rm SiO})$ & [cm$^{-3}$] & $1.3\times 10^5$  \\
   $n({\rm N}_2)$ & [cm$^{-3}$] & $5.6\times 10^6$  \\
\hline
   $D$ & [kpc] & 0.5  \\
   $\rho$ & & 0.1   \\
   $T$ & [K] &  2845  \\
   $n(\rm H)$ & [cm$^{-3}$] & $1.1\times 10^{12}$  \\
   $R$ & [au] & 2.8  \\
   $\tau_{\ell}$(232\,GHz) & & 1.2 \\
   $\tau_{\rm c}$(232\,GHz) & & 1.3  \\
   $n(e)$ & [cm$^{-3}$] & $3.9\times 10^7$  \\
   $M_{\rm gas}$ & [M$_{\odot}$] & $1.2\times 10^{-4}$  \\
   $n({\rm H}_2)$ & [cm$^{-3}$] & $5.2\times 10^7$  \\
   $n({\rm CO})$ & [cm$^{-3}$] & $3.0\times 10^8$  \\
   $n({\rm OH})$ & [cm$^{-3}$] & $3.7\times 10^4$  \\
   $n({\rm SiO})$ & [cm$^{-3}$] & $3.1\times 10^5$  \\
   $n({\rm N}_2)$ & [cm$^{-3}$] & $2.4\times 10^7$  \\
\hline
   $D$ & [kpc] & 2.0  \\
   $\rho$ & & 0.1   \\
   $T$ & [K] &  2770  \\
   $n(\rm H)$ & [cm$^{-3}$] & $6.0\times 10^{11}$  \\
   $R$ & [au] & 11  \\
   $\tau_{\ell}$(232\,GHz) & & 1.3 \\
   $\tau_{\rm c}$(232\,GHz) & & 1.3  \\
   $n(e)$ & [cm$^{-3}$] & $2.0\times 10^7$  \\
   $M_{\rm gas}$ & [M$_{\odot}$] & $4.2\times 10^{-3}$  \\
   $n({\rm H}_2)$ & [cm$^{-3}$] & $2.5\times 10^7$  \\
   $n({\rm CO})$ & [cm$^{-3}$] & $1.6\times 10^8$  \\
   $n({\rm OH})$ & [cm$^{-3}$] & $1.8\times 10^4$  \\
   $n({\rm SiO})$ & [cm$^{-3}$] & $2.3\times 10^5$  \\
   $n({\rm N}_2)$ & [cm$^{-3}$] & $1.4\times 10^7$  \\
\hline
\end{tabular}
\label{t:solutions2}
\tablefoot{(a) All of the above models satisfy the observed constraints at 232\,GHz:  $\theta$\,=\,11\,mas, $S_{\nu}(30\alpha)$\,=\,7.3\,mJy, and and $S_{\rm c}(232)$\,=\,7.5\,mJy.}
\end{table}

\begin{table}
 \caption{Models reflecting a larger range of parameter space, continued.}
   \begin{tabular}{lll}
\hline \hline   
   Property & & Primary solution  \\
\hline
   $D$ & [kpc] & 1.0  \\
   $\rho$ & & 0.1  \\
   $T$ & [K] &  2723  \\
   $n(\rm H)$ & [cm$^{-3}$] & $3.68\times 10^{11}$  \\
   $r$ & [au] & 8.5 \\
   $\tau_{\ell}$(232\,GHz) && 0.36 \\
   $\tau_c$(232\,GHz) & & 0.38  \\
   $n(e)$ & [cm$^{-3}$] & $1.2\times 10^7$  \\
   $M_{\rm gas}$ & [M$_{\odot}$] & $1.2\times 10^{-3}$ \\
   $n({\rm H}_2)$ & [cm$^{-3}$] & $1.3\times 10^7$ \\
   $n({\rm CO})$ & [cm$^{-3}$] & $9.9\times 10^7$  \\
   $n({\rm OH})$ & [cm$^{-3}$] & $9.4\times 10^3$  \\
   $n({\rm SiO})$ & [cm$^{-3}$] & $1.6\times 10^5$  \\
   $n({\rm N}_2)$ & [cm$^{-3}$] & $9.1\times 10^6$  \\
\hline
   $D$ & [kpc] & 1.0  \\
   $\rho$ & & 2.0  \\
   $T$ & [K] &  2580  \\
   $n(\rm H)$ & [cm$^{-3}$] & $9.34\times 10^{10}$  \\
   $r$ & [au] & 8.5 \\
   $\tau_{\ell}$(232\,GHz) && 0.39 \\
   $\tau_c$(232\,GHz) & & 0.40  \\
   $n(e)$ & [cm$^{-3}$] & $2.5\times 10^6$  \\
   $M_{\rm gas}$ & [M$_{\odot}$] & $6.8\times 10^{-4}$ \\
   $n({\rm H}_2)$ & [cm$^{-3}$] & $2.5\times 10^6$ \\
   $n({\rm CO})$ & [cm$^{-3}$] & $2.5\times 10^7$  \\
   $n({\rm OH})$ & [cm$^{-3}$] & $1.7\times 10^3$  \\
   $n({\rm SiO})$ & [cm$^{-3}$] & $7.1\times 10^4$  \\
   $n({\rm N}_2)$ & [cm$^{-3}$] & $2.6\times 10^6$  \\
\hline
\end{tabular}
\label{t:solutions2}
\tablefoot{(a) All of the above models satisfy the constraints at 232\,GHz:  $\theta$\,=\,17\,mas, $S_{\nu}(30\alpha)$\,=\,7.3\,mJy, and and $S_{\rm c}(232)$\,=\,7.5\,mJy.}
\end{table}

%
%
%
\section{Optical lines}
\label{a:opt_lines}

Table~\ref{t:opt_lines} shows a comparison with integrated line fluxes predicted by the reference model, including a correction for extinction. The simple equilibrium model was not tuned in any way to match the optical spectrum. The model nicely reproduces the integrated fluxes of the two Ca\,II forbidden lines. It under-predicts the 6300 and 6363\,\AA\ lines of O\,I by a factor of six and the two C\,I lines by a much larger factor. However, the amounts of atomic O and C in the gas are very sensitive to the abundance of CO, which in turn depends quadratically on density and exponentially on temperature in equilibrium. In the model, all of the tabulated lines have a peak optical depth $\tau$\,$<$\,0.7. Weak and forbidden lines were selected for a preliminary analysis because lines of higher optical depth would require a much more sophisticated treatment of radiative transfer through an extended atmosphere. Such analysis is far beyond the scope of the present work. Even so, it would be interesting for future work to consider that the RTL region can explain the fluxes of prominent forbidden lines in the visible spectrum within an order of magnitude.

\begin{table*}
\caption{Weak and forbidden lines at visible wavelengths}
\begin{tabular}{lccccccc}
\hline \hline \\
Species    &  $\lambda_0$\,$^a$  & $\upsilon_{\rm LSR}$   & $\Delta \upsilon$\,$^b$   & $S_{\nu,{\rm obs}}$   & $\int S_{\nu,{\rm obs}} {\rm d}\nu$   & $\int S_{\nu,{\rm model}} {\rm d}\nu$\,$^c$                       & $E_{\rm u}$\,$^d$ \\
           & [\AA]       & [km\,s$^{-1}$]         & [km\,s$^{-1}$]      & [Jy]        & [$10^{-15}$\,W\,m$^{-2}$]   & [$10^{-15}$\,W\,m$^{-2}$]   & [cm$^{-1}$] \\
 \hline \\
O\,I       &  5577.339     &  41.3(0.5)             & 20.7(1.1)           & \phantom{0}0.81(.04)  & 0.32(.05)         &  0.0017                     & 33793 \\
O\,I       &  6300.304     &  43.4(0.2)             & 24.0(0.5)           & \phantom{0}5.25(.08)  & 2.13(.05)         &  0.36\phantom{00}           & 15868 \\
O\,I       &  6363.776     &  41.9(0.3)             & 20.2(0.6)           & \phantom{0}2.02(.05)  & 0.68(.03)         &  0.12\phantom{00}           & 15868 \\
Ca\,II     &  7291.469     &  41.5(0.2)             & 14.7(0.4)           & 23.40(.40)             & 5.03(.17)        &  6.6\phantom{000}           & 13711 \\
Ca\,II     &  7323.887     &  41.4(0.1)             & 17.5(0.2)           & 18.90(.20)             & 4.79(.07)        &  5.0\phantom{000}           & 13650 \\
Fe\,II     &  7452.561     &  41.4(0.4)             & 15.2(0.8)           & \phantom{0}1.47(.07)  & 0.32(.02)         &  0.049\phantom{0}           & 15845 \\
Fe\,I      &  7723.207     &  41.4(0.7)             & 12.7(1.6)           & \phantom{0}0.60(.07)  & 0.11(.02)         &  5.1\phantom{000}           & 31323 \\
C\,I       &  8727.126     &  43.1(0.2)             & 16.3(0.4)           & \phantom{0}6.50(.13)  & 1.29(.04)         & $<0.001$                    & 21648 \\
C\,I       &  9850.250     &  39.8(0.8)             & 21.0(1.9)           & \phantom{0}0.96(.07)  & 0.22(.01)         & $<0.001$                    & 10193 \\
 \\
\hline \\
\end{tabular}
\label{t:opt_lines}
\tablefoot{(a) Wavelength in standard air. (b) FWHM of the Gaussian fit to the line profile.  (c) Integrated flux in the reference model, corrected for extinction at $A_{\rm V}$\,=\,0.99 magnitudes (estimated by comparing the observed flux density at 730\,nm with a Kurucz's  Atlas 9 grid model for $\log g$\,=\,1.5 and $T_{\rm eff}$\,=\,8500\,K at a distance of 1\,kpc). (d) Energy of the upper state of the transition. References: entries in column 2 come from the NIST Atomic Spectroscopy Database. The numbers in parentheses (\dots) are the root-mean-square uncertainties derived from the Gaussian fits.}
\end{table*}

\end{appendix}

\end{document}